\documentclass{jfm}
\pdfoutput=1 
\usepackage{graphicx}
\usepackage{mathtools}
\usepackage{units}
\usepackage{xcolor}
\usepackage{amssymb}
\usepackage{amsmath,mathrsfs}
\usepackage{amsmath,amssymb,amsfonts,bbm}
\usepackage[bbgreekl]{mathbbol}
\usepackage{mathrsfs}
 \usepackage{overpic} 
 \usepackage{hhline}
\usepackage{multirow}
\usepackage{array}
\usepackage{subfig}
\usepackage{wrapfig}
\usepackage{color}
\usepackage{epstopdf}
\usepackage{epsfig}
\usepackage{oplotsymbl}
\usepackage{tikz}
\usepackage{standalone}
\usepackage{pgfplots}
\usepackage{natbib}

\definecolor{darkcyan}{rgb}{0.0, 0.55, 0.55}

\shorttitle{Path instability of deformable bubbles rising in a viscous liquid}
\shortauthor{P. Bonnefis, J. Sierra-Ausin, D. Fabre \& J. Magnaudet}

\title{Path instability of deformable bubbles rising in Newtonian liquids: A linear study}

\author{
Paul Bonnefis, 
Javier Sierra-Ausin,
David Fabre
\&
Jacques Magnaudet 
  \corresp{\email{jmagnaud@imft.fr}}
 }

\affiliation{\aff{1}Universit\'e de Toulouse; INPT, UPS; IMFT (Institut de M\'ecanique des Fluides de Toulouse); All\'ee Camille Soula, F-31400 Toulouse, France}

\begin{document}
\maketitle

\abstract{
The first stages of the path instability phenomenon affecting the buoyancy-driven motion of gas bubbles rising in weakly or moderately viscous liquids are examined using a recently developed numerical approach designed to assess the global linear stability of incompressible flows involving freely-evolving interfaces. Predictions for the critical bubble size and frequency of the most unstable mode are found to agree well with reference data obtained in ultrapure water and in several silicone oils. By varying the bubble size, stability diagrams are built for several specific fluids, revealing three distinct regimes with different bifurcation sequences. The spatial structure of the unstable modes is analysed, together with the variations of the bubble shape, position and orientation. For this purpose, displacements of the bubble surface are split into rigid-body components and volume-preserving deformations, allowing us to determine how the relative magnitude of the latter varies with the fluid properties and bubble size. Predictions obtained with freely-deformable bubbles are compared with those found by maintaining the bubble shape determined in the base state frozen during the stability analysis. This comparison reveals that deformations leave the phenomenology of the first bifurcations unchanged in low-viscosity fluids, especially water. {\color{black}{Hence, in such fluids, bubbles behave essentially as freely-moving rigid bodies submitted to constant-force and zero-torque constraints, at the surface of which the fluid obeys a shear-free condition.}} In contrast, deformations change the nature of the primary bifurcation in oils slightly more viscous than water, whereas, somewhat surprisingly, they leave the near-threshold phenomenology unchanged in more viscous oils. 
}
\section{Introduction}
\label{sec:intro}
A great deal of effort has been invested since the middle of the last century to elucidate the origin and predict the characteristics of the intriguing path instability phenomenon affecting the buoyancy-driven motion of millimeter-sized gas bubbles rising in weakly or moderately viscous liquids. Attempts were mostly experimental during the first decades, before numerical simulation techniques became mature enough to shed some new light on this puzzling phenomenon. Among them, linear stability analyses have kept pace of advances in the computational techniques required to determine efficiently the discretized form of the stationary solutions of the Navier-Stokes equations subjected to appropriate boundary conditions, and to solve subsequently the large-size generalized eigenvalues problems {\color{black}{giving access to the possibly linearly unstable eigenmodes.}} \\
\indent Early studies considered bubbles with a frozen shape held fixed in a uniform stream. Assuming an oblate spheroidal shape with an arbitrary aspect ratio (the ratio of the major and minor axes lengths, hereinafter denoted as $\chi$), with the short axis of the spheroid aligned with the incoming flow, conditions under which the axisymmetric wake becomes unstable, the nature of the corresponding first bifurcations and the structure of the associated unstable modes could be explored \citep{Yang2007,Tchoufag2013}. It was eventually concluded that wake instability occurs within a finite range of Reynolds number when $\chi\gtrsim2.21$, in agreement with predictions provided by the numerical solution of the full  Navier-Stokes equations in the same configuration \citep{Magnaudet2007}. The first bifurcation is stationary, i.e. the corresponding eigenvalue is real. Beyond this bifurcation, the stationary wake exhibits a pair of counter-rotating trailing vortices which are responsible for a nonzero lift force. For larger aspect ratios, increasing the Reynolds number beyond the primary threshold while keeping $\chi$ fixed or \textit{vice versa}, a second bifurcation of Hopf type takes place. The  wake keeps the previous planar symmetry unchanged beyond the corresponding threshold but vortices are periodically shed downstream, resulting in periodic oscillations of the lift force about a nonzero mean. This transition scenario, with the axisymmetric wake becoming unstable through a stationary bifurcation and the resulting non-axisymmetric stationary wake undergoing a Hopf bifurcation, is similar to that obeyed by the wake of rigid spheres and discs \citep{Natarajan1993}. \\
\indent Compared with the actual problem of buoyancy-driven rising bubbles, the configuration contemplated in the above studies is purely academic, as the bubble is held fixed despite the existence of a nonzero lift force. An important step towards physically realistic conditions was achieved during the last decade, starting with global linear stability studies of the flow past freely moving buoyancy/gravity-driven rigid bodies with various shapes. In this situation, the body motion is subjected to two additional constraints, since the hydrodynamic force must balance the net body weight at all times (implying that the horizontal force remains null whatever the body orientation and velocity), and the hydrodynamic torque must also remain zero, provided that the body density is uniform. Therefore, disturbances in the body velocity or orientation modify the velocity and pressure distributions in the surrounding fluid, which in turn induces changes in the force and torque acting on the body that must respect the above contraints. Two-dimensional plates and rods were considered by \cite{Fabre2011} and \cite{Assemat2012}. The former study focused on `heavy' bodies with a density much larger than that of the fluid. An asymptotic approach could then be employed and revealed the existence of four `aerodynamic' modes distinct from the `fluid' (or `wake') modes associated with wake instability (qualitatively similar modes are encountered in flight dynamics, and those evidenced in the above studies were coined after them). Two of these aerodynamic modes have real negative eigenvalues, i.e. they are always stable. In contrast, the other two are associated with a pair of complex conjugate eigenvalues, hence with a Hopf bifurcation. The corresponding oscillations are slow compared with those typical of vortex shedding in two-dimensional wakes, and their frequency varies as the inverse square root of the body-to-fluid density ratio, $m^*$, thus becoming vanishingly small for very heavy bodies. Existence of these modes was confirmed for arbitrary $m^*$ in the global linear stability analysis (hereinafter abbreviated as GLSA) performed numerically by \cite{Assemat2012}, and the corresponding thresholds were compared with those of the fluid modes for thin plates and square rods.
Increasing the Reynolds number, it was observed that, for infinitely thin plates, the primary wake mode always becomes unstable first, no matter what the value of $m^*$ is. The same conclusion was found to hold for square rods as long as the density ratio exceeds $1.22$. In contrast, for \text{lower} $m^*$, the aerodynamic oscillatory mode is destabilized first and exhibits a frequency (corresponding to a fluttering of the falling/rising rod about the vertical) smaller than that of the wake mode by about a factor of two. Similar conclusions were previously obtained through fully-resolved numerical simulations  by \cite{Alben2008} with light two-dimensional ellipses, and the critical Reynolds number at which the fluttering motion sets in was found to be approximately half that corresponding to the onset of wake instability. The two-dimensional GLSA approach of \cite{Assemat2012} was extended to axisymmetric bodies by \cite{Tchoufag2014a} who considered the stability of rising/falling discs and short cylinders. Whatever the body aspect ratio, they found the aerodynamic low-frequency mode to be always the most unstable \text{beyond} a critical $\chi$-dependent $m^*$ (note the difference with two-dimensional square rods for which the same happens below a critical $m^*$). For lower $m^*$, the most unstable mode is either the one associated with the wake (here characterized by weak deviations of the path in the presence of sustained fluid oscillations at the back of the body), or a stationary mode in which the body follows an inclined path and its wake is made of a pair of steady counter-rotating streamwise vortices. Based on these findings, \cite{Tchoufag2014a} concluded that the first deviation of the path of such freely moving axisymmetric bodies cannot in general be predicted by considering the instability of the sole wake. Rather, the instability of their path results under most conditions from the intrinsic coupling between the body and fluid implied by the constant-force and zero-torque conditions.\\
\indent \cite{Tchoufag2014b} applied the GLSA approach to freely rising spheroidal bubbles with a prescribed oblateness. They observed that the vertical path becomes unstable within a finite range of Reynolds number when the bubble oblateness exceeds $2.15$, which is slightly lower than the wake instability threshold determined for a fixed bubble \citep{Magnaudet2007,Tchoufag2013}. The unstable Reynolds number range widens as $\chi$ increases. For $2.15<\chi<2.25$, increasing the Reynolds number, i.e. the bubble size, while maintaining the oblateness fixed, the path first becomes unstable within a narrow Reynolds number range through a low-frequency mode similar to the aerodynamic mode encountered with heavy short cylinders and discs. 
Increasing the Reynolds number further, a stationary mode yielding an inclined path and growing much faster than the previous low-frequency mode becomes dominant. This situation prevails within an intermediate Reynolds number range, beyond which the low-frequency mode becomes dominant again. 
 Last, the system restabilises beyond a fourth critical Reynolds number. For instance, the path of a bubble with $\chi=2.22$ becomes oscillatory through a low-frequency mode in the two separate ranges $310\lesssim Re\lesssim340$ and $450\lesssim Re\lesssim980$, while it takes a nonzero stationary inclination with respect to the vertical in the intermediate range $340\lesssim Re\lesssim450$, the Reynolds number $Re$ being based on the bubble rise speed and equivalent diameter. Obviously, the description provided by this simplified model suffers from the fact that the bubble shape is not allowed to vary with the Reynolds number, in contrast to what happens under real conditions. Nevertheless, as will become apparent later, the first half of the above scenario, i.e. the succession of a low-frequency mode becoming unstable first and a stationary mode taking over it, is relevant to the case of deformable bubbles rising in low-viscosity high-surface-tension fluids, especially water. \\
\indent A somewhat more sophisticated track was followed by \cite{Cano2016}. Rather than prescribing arbitrarily an oblate spheroidal shape, they computed the actual bubble shape and rise speed corresponding to the stationary axisymmetric base state, using the open software \textit{Gerris} \citep{Popinet2003,Popinet2007}. Then, they introduced this shape and speed into the GLSA solver of \cite{Tchoufag2014b}, keeping the shape frozen despite the possible development of flow and path instabilities. With this approach, they could determine for a wide variety of Newtonian fluids (characterized by their density, viscosity and surface tension), the critical Reynolds number, i.e. the critical bubble size, beyond which the vertical path of a bubble with a nearly realistic shape becomes unstable. They could also compare these predictions with those corresponding to `pure' wake instability, obtained by holding the same bubble fixed in a uniform stream. For low-viscosity high-surface-tension fluids, they found that, in agreement with the conclusions of \cite{Tchoufag2014b}, the system first becomes unstable through a Hopf bifurcation leading to low-frequency path oscillations. At the corresponding Reynolds number, the wake of the same bubble held fixed is still stable, which proves that the mechanism responsible for path instability is intimately linked to the constant-force and torque-free conditions that constrain the dynamics of the coupled fluid-body system. For liquids with a viscosity $5$ to $10$ times larger than water and a surface tension $3-4$ times smaller, they observed that path instability first arises through an oscillatory mode with a frequency $4-5$ times higher than that found in low-viscosity high-surface tension liquids. However, the picture looked dramatically different in the intermediate range corresponding to `moderate' liquid viscosities and surface tensions. In particular, the instability thresholds of freely-moving and fixed bubbles were found to be identical in that range. Hence, the first unstable mode is stationary, similar to that observed for fixed spheroidal bubbles \citep{Yang2007,Magnaudet2007,Tchoufag2013}. When compared with reference experiments carried out in ultrapure water \citep{Duineveld1995, DeVries2002} or silicone oils \citep{Zenit2008, Sato2009}, these predictions experience mixed successes. In low-viscosity high-surface-tension fluids, they properly capture the low-frequency oscillatory nature of the incipient path instability, and slightly over-predict the size of the smallest bubble whose path becomes unstable. The same conclusion holds for the `high'-frequency path oscillations observed in oils $5-10$ times more viscous than water. In contrast, for oils with an intermediate viscosity, experimental observations provide evidence that the first instability of the system keeps a similar oscillatory nature, at odds with the above predictions. \\
\indent This blatant disagreement points to the role of transient bubble deformations not accounted for in the above studies. It provided one of the main motivations that decided us to undertake the development of a more general linear stability approach capable of dealing with free gas-liquid boundaries, thanks to which the fate of time-dependent deformations of the bubble-fluid interface may be predicted together with that of flow and path disturbances. {\color{black}{The corresponding method was successfully developed and validated by \cite{Bonnefis2019}}}; other groups have pursued similar objectives in parallel \citep{Zhou2017,Herrada2023}. {\color{black}{These efforts may be considered as the third and final step on the long route of the linear stability analysis of the flow past rising bubbles. To summarize, in the first step the bubble shape was frozen, its centroid was assumed to remain fixed and only the stability of the wake was assessed \citep{Yang2007,Tchoufag2013}. In the next step, the bubble shape was still frozen but the bubble centroid was allowed to move freely under the effect of buoyancy, being only constrained by the constant-force and zero-torque conditions \citep{Tchoufag2014b,Cano2016}. Most of the investigations performed in these first two steps assumed the bubble to keep an exact oblate spheroidal shape, except the last of them in which this shape was obtained as part of the stationary solution of the full Navier-Stokes equations. The time-dependent adjustment of the bubble shape to its possibly non-straight unsteady motion was the last ingredient missing to reach a fully realistic description of the bubble-flow interaction. This is what the third step achieved by \cite{Bonnefis2019} enables.}} The purpose of this paper is to present and discuss the predictions provided by this approach across a wide range of fluids, from low-viscosity high-surface-tension liquid metals to oils with a viscosity typically ten times larger and a surface tension four times less than those of water, and to compare them with reference data when available.  {\color{black}{Predictions obtained in the specific case of water were already summarized by \cite{Bonnefis2023} and are discussed in more detail here.}} The numerical approach, already described by \cite{Sierra2022}, is summarized in \S\,\ref{numer}. The characteristics of the base state are presented in \S\,\ref{base}, before the neutral curves resulting from the linear stability analysis are discussed in \S\,\ref{neutral}, and the nature and sequencing of the first unstable modes are examined in \S\,\ref{unstable}. Then, the relative magnitude of deformations of the bubble-fluid interface with respect to the lateral drift of the bubble centroid are discussed in \S\,\ref{deform}. Section \ref{conclu} summarizes the main findings of the study and discusses the respective roles of wake, bubble deformation and fluid-bubble hydrodynamic couplings in the different regimes encountered by varying the physical properties of the carrying liquid. 
\section{Problem formulation and numerical approach}
\label{numer}
We assume that a single gas bubble with volume $\mathcal{V}$ and equivalent diameter $D=(\frac{6}{\pi}\mathcal{V})^{1/3}$ rises through a large expanse of a Newtonian liquid with density $\rho$,  viscosity $\mu$ and surface tension $\gamma$ under the effect of gravity $g$. The liquid is at rest at infinity and the disturbance flow induced by the bubble ascent is assumed incompressible. The bubble is initially axisymmetric and is released with its symmetry axis aligned with gravity. Considering that the density and viscosity of the gas enclosed in the bubble are negligibly small compared with those of the surrounding liquid, the problem depends on two dimensionless parameters, namely the Bond number $Bo=\rho g D^2/\gamma$ and the Galilei number $Ga=\rho(gD^3)^{1/2}/\mu$. These characteristic numbers compare the gravitational force $\rho g D^3$ driving the bubble ascent with the capillary force $\gamma D$ and the viscous force $\mu(gD^3)^{1/2}$, respectively. Whatever the bubble size, a given liquid placed in a given gravitational environment is entirely characterized by the value of the so-called Morton number $Mo=Bo^3Ga^{-4}=g\mu^4/(\rho\gamma^3)$. By defining the visco-gravitational diffusion length scale $l_\mu=[\mu^2/(\rho^2g)]^{1/3}$ (the length over which viscosity diffuses a disturbance within a $[\mu/(\rho g^2)]^{1/3}$-long period of time) and the capillary length scale $\l_\gamma=[\gamma/(\rho g)]^{1/2}$, it is seen that $l_\mu/l_\gamma=Mo^{1/6}$. Hence, the Morton number characterizes the ratio of these two length scales. {\color{black}{The bubble diameter, $D$, the gravitational time, $(D/g)^{1/2}$, and their ratio, $(gD)^{1/2}$, will be used throughout the paper to make lengths, frequencies and velocities dimensionless. Once the bubble rise speed $u_b$ is known, the non-dimensional rise speed, $U=u_b/(gD)^{1/2}$ allows the bubble Reynolds number, $Re=Ga\,U$, and Weber number, $We=Bo\,U^2$, to be evaluated. }}\\
\begin{figure}
\vspace{5mm}
\centering
  \includegraphics[width=0.4\textwidth]{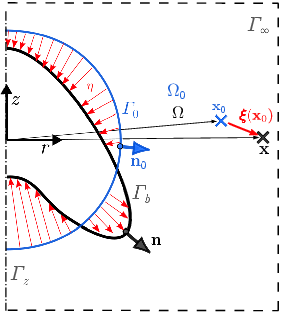}
  \vspace{0mm}\\
    \caption{{\color{black}{Flow domain and geometrical transformation involved in the L-ALE approach. The black line represents the current bubble-fluid interface $\Gamma_b(t)$ bounding internally the physical fluid domain $\Omega(t)$, while the blue line corresponds to the initial interface $\Gamma_{b0}$ bounding the reference domain $\Omega_0$. Both domains exhibit a rotational symmetry about the $\Gamma_z$ axis.}}}
\label{fig:sketch}
\end{figure}
\indent The stability of the coupled fluid-bubble motion is examined with the help of the Linearized Arbitrary Lagrangian-Eulerian approach, hereinafter abbreviated as L-ALE, developed by \cite{Bonnefis2019} and described by \cite{Sierra2022}. Here, we only summarize the main steps of this approach, referring readers to the above references for more details. The key idea is to solve the governing equations of the problem, which hold on a time-deforming domain $\Omega(t)$ since the bubble-fluid interface $\Gamma_b(t)$ experiences time-dependent deformations, on a prescribed reference domain $\Omega_0$ bounded internally by a fixed surface $\Gamma_0$ representing the reference position of the interface (figure \ref{fig:sketch}). In addition to $\Gamma_0$, $\Omega_0$ is bounded by a fixed external surface $\Gamma_\infty$ on which suitable far-field conditions are imposed, and $\Gamma_0$ and $\Gamma_\infty$ are connected by the vertical axis, $\Gamma_z$, about which the base state exhibits an axial symmetry. The L-ALE approach is grounded on the explicit building of the bijective mapping connecting the instantaneous position $\bf{x}$ of a given geometrical point in the deforming domain to its position $\bf{x}_0$ in $\Omega_0$. In other words, this approach relies on the determination of the infinitesimal displacement field $\boldsymbol{\xi}(\bf{x}_0)$ such that $\bf{x}=\bf{x}_0+\boldsymbol{\xi}(\bf{x}_0)$ everywhere in $\Omega_0\cup\Gamma_0\cup\Gamma_z\cup\Gamma_\infty$. On $\Gamma_0$, the unit normal of which is $\bf{n}_0$, the normal component of the displacement, $\boldsymbol{\xi}\cdot\bf{n}_0$, must equal the displacement of the bubble-fluid interface, $\eta$. In contrast,  $\boldsymbol{\xi}$ may be arbitrarily chosen anywhere else. To ensure a smooth distribution of the displacement field, we assume that $\boldsymbol{\xi}$ obeys the Cauchy-Navier equation of elastostatics with unit Lam\'e coefficients in $\Omega_0$, together with a suitable symmetry condition on $\Gamma_z$ and a Dirichlet condition $\boldsymbol\xi={\bf{0}}$ on $\Gamma_\infty$.  \\
\indent Knowing $\boldsymbol{\xi}(\bf{x}_0)$ and its gradients everywhere, the time and space derivatives involved in the governing equations written in $\Omega(t)$ may be expressed in $\Omega_0$ and linearized consistently with respect to $\boldsymbol{\xi}$, together with the variations of the unit normal and local mean curvature of $\Gamma_b(t)$. Governing equations in $\Omega_0$ were detailed in \cite{Sierra2022}. They consist of the continuity and Navier-Stokes equations, while the kinematic (no-penetration) and dynamic boundary conditions must be enforced on $\Gamma_0$. Owing to the negligible viscosity of the gas enclosed in the bubble, the tangential component of the dynamic boundary condition reduces to a shear-free condition that must be satisfied on $\Gamma_0$ by the gradients of the liquid velocity field, $\bf{u}$. Since the gas density is also negligible, the momentum balance implies that the pressure is uniform (but possibly time-dependent) within the bubble. Consequently, the normal component of the dynamic boundary condition on $\Gamma_0$ expresses the fact that the difference between the uniform pressure $p_b$ inside the bubble and the local pressure $p$ in the liquid (from which the hydrostatic component has been subtracted) balances the difference between the local capillary pressure and the normal viscous stress in the liquid. Last, two integral conditions have to be added to ensure that the volume of the bubble does not change over time, and that the vertical position of its geometrical centroid does not vary once the problem has been expressed in a reference frame translating with the bubble rise speed, $u_b$. \\
\indent  A key feature of the L-ALE approach is that the entire set of equations governing the evolution of the state vector ${\bf{q}}=[{\bf{u}},p,\eta,\boldsymbol\xi,u_b, p_b]^\text{T}$ ($^\text{T}$ denoting the transpose) is solved simultaneously (such an approach is frequently referred to as `monolithic'). Starting from an arbitrary initial solution ${\bf{q}}_0=[{\bf{u}}_0,p_0,0,{\bf{0}},u_{b0},p_{b0}]^\text{T}$ (here assumed to be axisymmetric) and dropping time derivatives in the Navier-Stokes equations and in the no-penetration condition, a Newton algorithm is used to  obtain the stationary base state through a series of global iterations in which the state vector and the reference domain are iteratively updated. For this purpose, we express the problem in weak form and make use of the finite element software FreeFem++ \citep{Hecht2012} to build and invert the various matrices involved. The volume fields $({\bf{u}},p,{\boldsymbol\xi})$ are discretized on the finite element basis suitable for each of them, while the normal displacement of the interface, $\eta$, is discretized on the local Fourier basis. Provided that ${\bf{q}}_0$ is chosen `not too far' from the stationary solution (i.e. within its basin of attraction), the above algorithm converges quadratically in a few iterations. Moreover, compared with time-marching algorithms, it has the decisive advantage that unstable steady solutions may also be captured, allowing the features of the corresponding branches (if any) of the bifurcation diagram to be studied in detail; see, e.g., \cite{Sierra2022}.\\
\indent Once the state vector and the shape of the bubble-fluid interface defining the axisymmetric base state have been obtained, the linear stability of this solution is assessed in the Galilean reference frame translating with the corresponding rise speed. For this purpose, we consider a cylindrical coordinate system $(r,\theta,z)$ whose axis $r=0$ and cross-sectional plane $z=0$ correspond to the symmetry axis and vertical position of the bubble geometrical centroid in the base state, respectively. Then we compute the complex eigenvalues $\lambda=\lambda_r+\text{i}\lambda_i$ associated with disturbances of the form ${\bf{q}}'(r,\theta,z,t)=[({\hat{\bf{u}}},\hat{p},\hat{\eta},\hat{\boldsymbol\xi})e^{\text{i}m\theta},\hat{p}_b]^\text{T}e^{\lambda t}$, with $m$ denoting the azimuthal wave number and $\theta$ the azimuthal angle, the hatted amplitudes $({\hat{\bf{u}}},\hat{p},\hat{\eta},\hat{\boldsymbol\xi})$ depending on both $r$ and $z$ (owing to the choice of the reference frame, $\hat{u}_b\equiv0$, so that the instantaneous position of the bubble centroid no longer necessarily coincides with that of the reference frame). In what follows, we shall be primarily interested in the family of non-axisymmetric modes $|m|=1$ which are those associated with the first deviations of the bubble path from the vertical, and in modes $m=0$ and $|m|=2$ associated with the oscillations of the bubble shape. The eigenvector ${\bf{q}}'$ is obtained $via$ the Newton algorithm  already employed to determine the base state. The set of equations to be solved is similar, except that the momentum equation and the no-penetration condition now involve  terms proportional to $\lambda$ since the sought solution is time-dependent; the reference state vector and domain to be considered are those corresponding to the base state. The complex eigenvalues are finally obtained using a Krylov-Shur projection method available in the SLEPc library. \\
\indent Obviously, the accuracy with which the base state and the eigenvalues are determined depends critically on the finite element triangulation of $\Omega_0$, especially along the bubble-fluid interface $\Gamma_0$ and within the boundary layer and wake regions. The position of the fictitious outer boundary $\Gamma_\infty$ with respect to the bubble is also important, the computational domain having to be large enough to avoid spurious confinement effects. It must also be kept in mind that, in low-Morton-number fluids such as water ($Mo\approx2.54\times10^{-11}$ under standard conditions) and even more in liquid metals, such as Galinstan  ($Mo\approx1.4\times10^{-13}$), the onset of path instability takes place in regimes where the bubble Reynolds number $Re=\rho u_bD/\mu$ is of the order of $10^3$ ($\approx670$ in water, $\approx2100$ in Galinstan, see figure \ref{fig:neutralcurve} below). To ensure that the flow is still fully resolved in the thin boundary layers encountered in such regimes, we design an initial grid in which $18$ nodes are distributed across a $5\times Re^{-1/2}$-thick layer surrounding the bubble. 
 Along the bubble surface, the number of nodes $N_D$ over a $D$-long arc length is approximately $7.2\times Re^{1/2}$, with at least $N_D=64$ when $Re$ becomes less than $80$. Since $u_b$, hence $Re$, is not known \textit{a priori}, once an approximate steady state has been computed on the initial grid, a new grid is generated following an adaptive mesh refinement procedure. This new grid is designed based on the previous steady-state solution, with some constraints ensuring that the boundary layer remains sufficiently resolved. This procedure is repeated a few times until the solution becomes grid-independent. Extensive tests were carried out to evaluate the sensitivity of the base state (appreciated for instance through the variations of $Re$, $\chi$, $\lambda_r$ and $\lambda_i$) to factors such as the domain size, distance from the bubble to the downstream boundary and density of nodes along the bubble surface. For instance, in the case of a bubble with $Bo=0.6$ rising in water, it was found that relative variations of these indicators for the non-axisymmetric modes $m=\pm1$ are all less than $0.1\%$ provided that $N_D\gtrsim70$ and the distance from the bubble centroid to the downstream end of the domain is at least $30D$, the upstream end and the lateral surface being both located $15D$ away from this centroid \citep{Bonnefis2019}.

\section{Base state}
\label{base}
\subsection{Bubble shapes}
\label{shapes}
\begin{table}
\centering
\begin{tabular}{c c c c c c}
\hline
 Liquid   & $\rho$ (kg\,m$^{-3}$) & $\mu$ (mPa\,s) & $\gamma$ (mN\,m$^{-1}$) & $Mo=g\mu^4/(\rho\gamma^3)$               \\ \hline
 Galinstan         & $6440$            & $2.40$               & $718$             & $1.4\times10^{-13} $   \\
 Water ($28^\circ \text{C}$, 1 atm) & $996$            & $0.80$               & $71.0$           & $1.13\times10^{-11} $                \\
 Water ($20^\circ \text{C}$, 1 atm) & $1000$            & $1.00$               & $72.8$           & $2.54\times10^{-11} $     \\
 DMS-T00           & $761$             & $0.49$               & $15.9$            & $1.8\times10^{-10} $                   \\
 DMS-T02           & $873$             & $1.75$               & $18.7$            & $1.6\times10^{-8} $                   \\
 DMS-T05           & $918$             & $4.59$               & $19.7$            & $6.2\times10^{-7} $                  \\
 DMS-T11           & $935$             & $9.35$               & $20.1$            & $9.9\times10^{-6} $                  \\ \hline
 \end{tabular}
\caption{Physical properties of some specific fluids considered in this study. Note that the surface tension of Galinstan may vary from $535$ to $718\,$mN\,m$^{-1}$, depending on the exact alloy composition.}
\label{tab1}
\end{table}
\begin{figure}
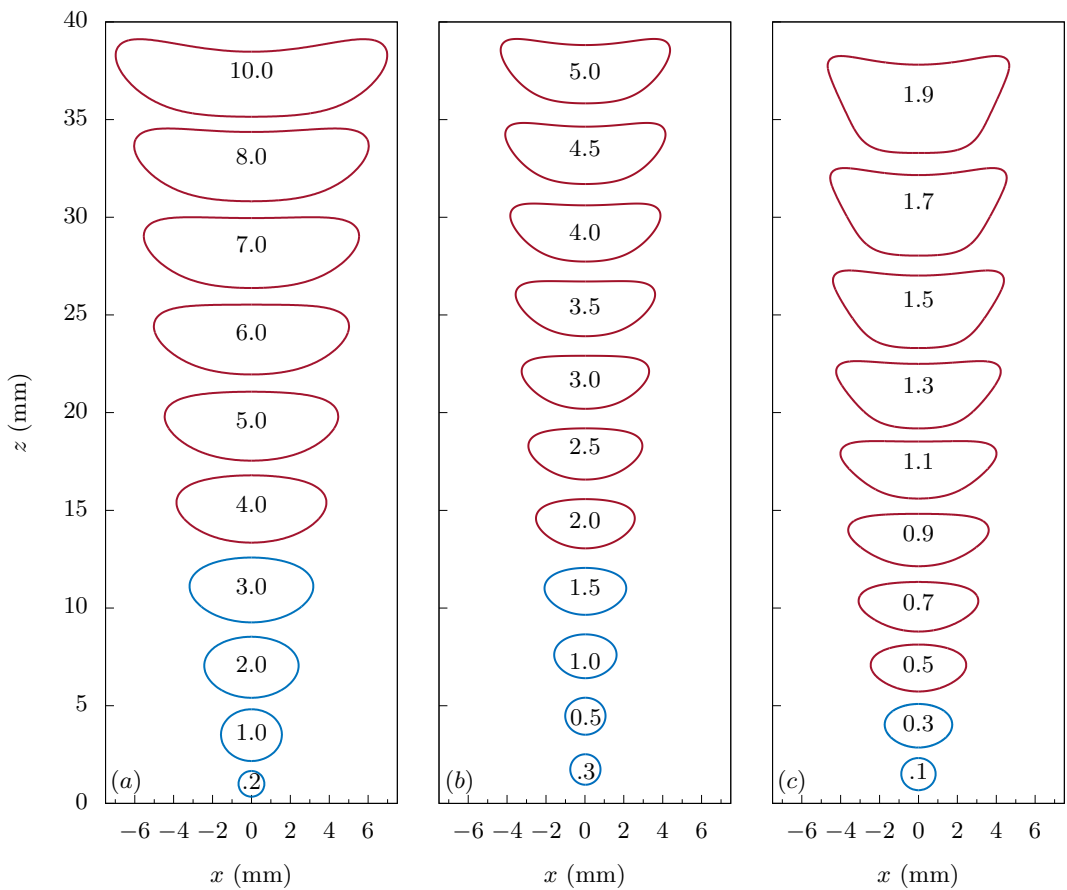

\vspace{5mm}
\hspace{-15mm}
	{\input{Figures/Shapes_DMST05.tex}\input{Figures/Shapes_DMST02.tex}\input{Figures/Shapes_eau20.tex}}		
\vspace{-18mm}\\
\hspace{5mm}${}$ \hspace{5mm}$(a)$\hspace{40mm}$(b)$\hspace{40mm}$(c)$\\
\vspace{11mm}
\caption{Equilibrium shapes of bubbles with {\color{black}{various Bond numbers}} in three different fluids; the Bond number is specified within each contour. {\color{black}{Contours represent different bubbles;}} their centroids are arbitrarily shifted in the vertical direction for readability, {\color{black}{making the local value of $z$ irrelevant}}. Blue and red contours refer to stable and unstable configurations, respectively. $(a)$: silicone oil DMS-T05 ($Mo\!=6.2\!\times 10^{-7}$); $(b)$: silicone oil DMS-T02  ($Mo=1.6\times10^{-8}$); $(c)$: water at $20^\circ \text{C}$ ($Mo\!=\!2.54\times 10^{-11}$). }
\label{fig:shapes}
\end{figure}
Figure \ref{fig:shapes} shows how the equilibrium shape of the bubble changes with the Bond number in three different fluids. {\color{black}{As the Morton number keeps a constant value in each fluid, increasing $Bo$ (or $Ga$) in a given series amounts to increasing the bubble size:}} since $Bo\propto D^2$, the volume of the bubble grows as $Bo^{3/2}$. In each series, starting from nearly spherical shapes at low Bond number, bubbles take oblate spheroidal shapes when finite-$Bo$ effects become significant. The figure makes it clear that the properties of the carrying fluid have a major influence on the range  of sizes in which this geometric approximation is valid: while the shape of a bubble with $Bo=3$ is still close to a spheroid in silicone oil DMS-T05 ($Mo= 6.2\times10^{-7}$, see table \ref{tab1} for the physical properties of the various fluids discussed below), a bubble with $Bo=0.5$ is seen to already exhibit a significant fore-aft asymmetry in water. Beyond that range, the front part of the bubble flattens increasingly as its size increases, while its rear part becomes more and more rounded. As indicated by the change in colour of the bubble contour in the figure, the transition between these two `regimes' corresponds approximately to the onset of path instability. Although they do not rise in a straight line, bubbles with a mildly convex front and a pronounced rounded rear are routinely observed in experiments. In contrast, bubbles with an almost flat or even concave front such as those shown in the figure for $Bo\gtrsim5$ in DMS-T05 or $Bo\gtrsim0.7$ in water are not. The obvious reason is that such bubbles actually experience large shape oscillations, so that the present stationary solution makes little sense in these regimes.
\begin{figure}
\vspace{5mm}
\hspace{-5mm}
  \input{Figures/Re_vs_Bo_latex.tex}
  \vspace{-8mm}\\
    \caption{Rise Reynolds number vs the Bond number in several fluids. Solid lines: present predictions; symbols: experimental data, with $\textcolor{black}{\squaddot}$\,: ultrapure water at $20^\circ$C {\color{black}{($Mo\!=\!2.54\times 10^{-11}$)}} \citep{Duineveld1995}, $\textcolor{orange}{\scross}$\,: silicone oil DMS-T00  {\color{black}{($Mo=1.8\times10^{-10}$)}}, $\textcolor{yellow}{\circletdot}$\,: DMS-T02  {\color{black}{($Mo=1.6\times10^{-8}$)}}, $\textcolor{green}{\rhombusdot}$\,: DMS-T05  {\color{black}{($Mo\!=6.2\!\times 10^{-7}$)}}, and $\textcolor{purple}{\trianglepbdot}$\,: DMS-T11  {\color{black}{($Mo=9.9\times10^{-6}$)}} (data for all four series taken from \cite{Zenit2008}). Red bullets mark the onset of path instability detected experimentally in each series. }
\label{fig:Re}
\end{figure}
\subsection{Rise speed}
\label{speeds}
\indent Variations of the equilibrium rise speed with the bubble size and liquid properties are displayed in figure \ref{fig:Re} in the form of a $Re(Bo)$ diagram. Comparison with experimental data reveals an excellent agreement in most fluids, especially in the most challenging case of water. A consistent $10\%$ over-estimate may be noticed in the least viscous silicone oil, DMS-T00, presumably because of some variation in the actual viscosity or surface tension of the sample used in the experiments with respect to the reference values provided by the manufacturer \citep{Zenit2008}. In each series, the location of the incipient path instability detected in the experiments is specified in the figure. The threshold Reynolds number is seen to vary by nearly one order of magnitude from water to the most viscous oil, DMS-T11, $9.4$ times more viscous than water. As the three intermediate series make clear, the predicted Reynolds number departs from that determined in the experiments beyond the threshold. This is no surprise, since part of the potential energy of zigzagging or spiralling bubbles is transferred to the radial and azimuthal fluid motions, yielding a decrease in the rise speed compared with the rectilinear regime.\\
{\color{black}{\indent Data reported in figure \ref{fig:Re} may also be used to shed some light on the influence of the bubble shape and properties of the carrying fluid on the bubble rise speed. To this end, the above numerical predictions for the rise Reynolds number are replotted against the Galilei number in figure \ref{fig:Ga}. This figure reveals the succession of two distinct behaviours in each fluid. First, at low enough $Ga$, all data collapse on a master curve following the power law $Re\propto Ga^{5/3}$. This behaviour corresponds to the regime in which the bubble rise speed does not depend on surface tension, hence on the Morton number. In this regime, the bubble shape remains close to a sphere, i.e. its aspect ratio $\chi$ is close to unity. For low enough $We$, departures from sphericity are known to increase linearly with the Weber number according to the law $\chi=1+\frac{9}{64}We$  \citep{Moore1965}. 
Since $We=Re^2Mo^{1/3}Ga^{-2/3}$, the lower $Mo$ the wider the $Re$ range pertaining to this first regime, which extends up to $Re\approx200$ in water. 
Bubbles become flatter as $We$, hence $Re$, increases, opposing a larger resistance to the fluid. This makes their rise speed grow more slowly with $Ga$ when the Weber number becomes of order unity. Since their frontal area depends on $We$ which itself depends on $Mo$, the Reynolds number is no longer independent of the Morton number in this second regime. 
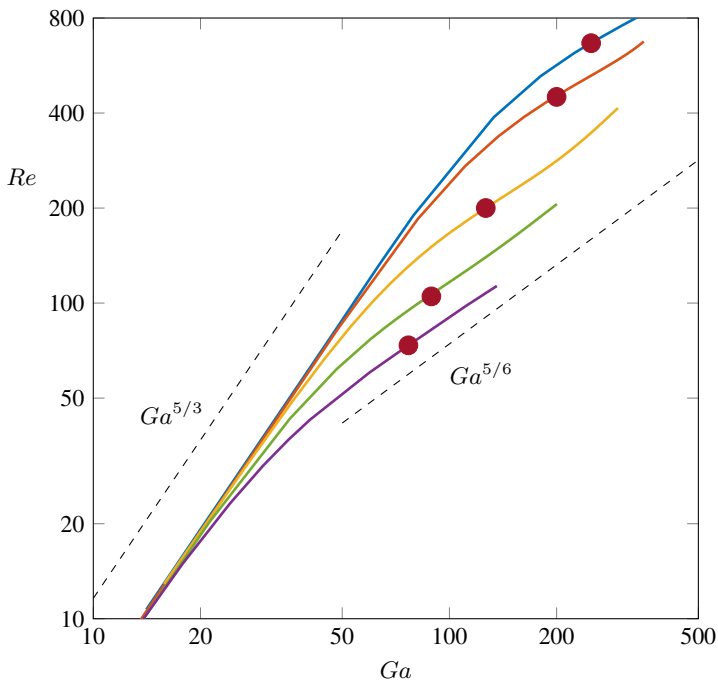
\begin{figure}
\vspace{5mm}
\centering
 {\color{black}{
  {\definecolor{mycolor1}{rgb}{0.00000,0.44700,0.74100}%
\definecolor{mycolor2}{rgb}{0.85000,0.32500,0.09800}%
\definecolor{mycolor3}{rgb}{0.92900,0.69400,0.12500}%
\definecolor{mycolor4}{rgb}{0.46600,0.67400,0.18800}%
\definecolor{mycolor5}{rgb}{0.49400,0.18400,0.55600}%
\definecolor{mycolor6}{rgb}{0.63922,0.07843,0.16863}%
\definecolor{mycolor7}{rgb}{0.30100,0.74500,0.93300}%
\definecolor{mycolor8}{rgb}{0.63500,0.07800,0.18400}%
\begin{tikzpicture}

\begin{axis}[%
width=8cm,
height=7.909cm,
at={(0cm,0cm)},
scale only axis,
xmode=log,
xmin=10,
xmax=500,
xminorticks=true,
xlabel style={font=\color{white!15!black}},
xlabel={$Ga$},
ymode=log,
ymin=10,
ymax=800,
yminorticks=true,
ylabel={$Re$},
axis background/.style={fill=white},
ylabel near ticks,
xlabel near ticks,
ylabel style={font=\small\fontfamily{ptm}\selectfont,anchor=base,rotate=-90,xshift=-0.25cm,yshift=1.75cm},
xlabel style={font=\small\fontfamily{ptm}\selectfont},
legend style={font=\small\fontfamily{ptm}\selectfont},
ticklabel style={font=\small\fontfamily{ptm}\selectfont},
clip=true,
xtick = {5,10,20,50,100,200,500,1000},
xticklabels = {5,10,20,50,100,200,500,1000},
ytick = {1,2,5,10,20,50,100,200,400,800},
yticklabels = {1,2,5,10,20,50,100,200,400,800},
major tick length = 0.15 cm,
minor tick length = 0.075 cm
]
\addplot [color=mycolor1, line width=1.0pt, forget plot]
  table[row sep=crcr]{%
14.0857933420126	10.6909\\
47.0997548469334	79.9747\\
63.8389607745408	133.441\\
79.2117181661414	189.442\\
133.218617540215	388.09\\
180.562959693629	524.183\\
224.043690219664	619.436\\
250.023198276393	667.356\\
303.674572184461	755.104\\
313.113349368623	769.4\\
322.457057351123	783.346\\
331.714032365315	796.995\\
340.886297264671	810.381\\
349.985279275649	823.56\\
358.994271356662	836.518\\
367.937312857198	849.347\\
376.798038775467	862.041\\
385.592151247117	874.689\\
394.326319405434	887.368\\
402.996273557107	900.141\\
411.596566363105	913.082\\
420.148676656526	926.354\\
428.635063393894	940.115\\
437.065287067749	954.748\\
445.43584158151	970.86\\
453.770417566148	989.436\\
462.03876655721	1011.55\\
470.265606418614	1037.71\\
478.449006668956	1064.49\\
486.586224691824	1088.47\\
494.675134690875	1109.97\\
502.712337024961	1129.78\\
510.711686196327	1148.37\\
518.671974680319	1166.16\\
526.589712928715	1183.29\\
534.464239950274	1199.48\\
542.3102372245	1214.6\\
550.109034716365	1228.87\\
557.87592860513	1242.57\\
565.610303342899	1255.55\\
};
\addplot [color=mycolor2, line width=1.0pt, forget plot]
  table[row sep=crcr]{%
351.382649557804	674.254\\
346.666048473351	665.842\\
341.927483249968	657.641\\
337.168348269188	649.647\\
332.382062516994	641.839\\
327.576928100457	634.222\\
322.747267412075	626.764\\
317.894294883639	619.45\\
313.01908978098	612.263\\
308.110182664024	605.159\\
303.181600496069	598.15\\
298.228594341123	591.209\\
293.240515886916	584.296\\
288.230964285082	577.427\\
283.189708993319	570.564\\
278.113299091808	563.687\\
273.014361419025	556.81\\
267.87843206191	549.75\\
262.70840913332	542.8\\
257.50612887382	535.785\\
252.268574203825	528.691\\
246.994172696756	521.501\\
241.684658813285	514.21\\
236.329924418689	506.783\\
230.941069936323	499.234\\
225.504772668128	491.517\\
220.028545391299	483.636\\
214.503693917191	475.554\\
208.930761827733	467.254\\
203.309886064885	458.716\\
197.635007847548	449.903\\
191.907472983204	440.793\\
186.11967933611	431.337\\
162.332379529097	389.288\\
137.316856014463	337.717\\
110.667730857369	271.515\\
81.6493178894077	185.118\\
48.5489748846013	82.6796\\
28.8687163978508	34.8667\\
8.63378396302043	4.66699\\
};
\addplot [color=mycolor3, line width=1.0pt, forget plot]
  table[row sep=crcr]{%
297.290166533931	414.837\\
288.329260777798	401.613\\
279.269899505984	388.318\\
270.122145346956	375.011\\
260.870695156163	361.718\\
251.482756610607	348.44\\
241.995351242586	335.366\\
232.379000772445	322.702\\
222.631174873125	310.02\\
212.733165472549	297.578\\
202.671790181263	285.35\\
192.455525806856	273.408\\
182.049360837355	261.632\\
171.444550637919	249.897\\
160.613831887254	238.068\\
149.532775361823	225.969\\
138.171221673167	213.386\\
126.492292288652	200.048\\
114.438495221875	185.546\\
101.940743355645	169.325\\
98.741756671137	164.918\\
95.5018542575892	160.326\\
92.2253778679976	155.545\\
88.9163324361369	150.569\\
82.1631340939392	139.89\\
75.2161755837993	128.105\\
68.0478757719785	115.027\\
60.6189699784504	100.488\\
52.8708764246845	84.397\\
44.7235958975237	66.8838\\
36.0445368676352	48.429\\
26.5927676655069	29.976\\
15.8121048355461	12.8867\\
};
\addplot [color=mycolor4, line width=1.0pt, forget plot]
  table[row sep=crcr]{%
6.33743161144087	2.74084\\
21.1904797430109	20.2446\\
35.6377669642207	43.0171\\
48.3030494933371	61.8799\\
59.9340842355393	76.7592\\
65.4688049074461	83.0997\\
70.8535822544844	88.9029\\
76.1035823597249	94.2762\\
81.2353620004322	99.3107\\
86.2603261822223	104.073\\
91.1919364913219	108.626\\
96.034383836934	113.006\\
100.795656465137	117.248\\
105.484128919561	121.382\\
110.107897000134	125.432\\
114.661159961461	129.403\\
119.160373948231	133.324\\
127.990768854716	141.042\\
136.618625534754	148.615\\
145.075098525121	156.108\\
153.365450251168	163.519\\
161.508896224594	170.854\\
169.520283231397	178.108\\
177.406529128683	185.261\\
185.179194476224	192.302\\
192.837483063218	199.198\\
200.400395965858	205.943\\
};
\addplot [color=mycolor5, line width=1.0pt, forget plot]
  table[row sep=crcr]{%
3.17015388764809	0.775182\\
10.6004070712213	6.56143\\
17.8276435921171	14.9438\\
24.1636539084018	23.137\\
29.9823123206909	30.5472\\
35.4441945061018	37.0479\\
40.6378904616073	42.7399\\
59.6093857841043	60.224\\
76.7204642145416	73.4911\\
78.7664761057808	75.0019\\
80.7948834569829	76.4888\\
82.807583338349	77.955\\
84.8032565697701	79.3997\\
86.7822916707304	80.8242\\
88.7471934855614	82.2315\\
90.698795243604	83.6229\\
92.6340773814916	84.9955\\
94.5580696204871	86.3547\\
96.4678233885008	87.6975\\
98.3667656913075	89.0274\\
100.251878109732	90.3413\\
109.50979377901	96.7439\\
111.331592200976	97.9773\\
135.880941991424	113.309\\
};

\addplot [color=black, dashed, forget plot, domain=10:50 ]
{0.25*x^1.66666};

\addplot [color=black, dashed, forget plot, domain=50:500 ]
{1.6*x^0.83333};

\node[right, align=left, inner sep=0]
at (axis cs:13.5,45) {\small$Ga^{5/3}$};

\node[right, align=left, inner sep=0]
at (axis cs:100.0,60) {\small$Ga^{5/6}$};

\addplot [color=mycolor5, only marks, mark size=3.5pt, mark=*, mark options={solid, fill=mycolor6, draw=mycolor6}, forget plot]
  table[row sep=crcr]{%
250	665.902\\
200	450\\
126.4923	200.05\\
89	105\\
76.7205	73.4911\\
};

\end{axis}
\end{tikzpicture}%
}
  }}
  \vspace{0mm}\\
    \caption{{\color{black}{Numerical predictions for the rise Reynolds number, plotted vs the Galilei number. The colour code and the meaning of the red bullets are similar to those of figure \ref{fig:Re}.}} }
\label{fig:Ga}
\end{figure}
To further examine the two behaviours, on has to refer to the force balance on the bubble which, in non-dimensional form, reads $C_D(Re)Re^2=\frac{4}{3}Ga^2$, with $C_D(Re)$ the usual drag coefficient. The drag coefficient of a spherical bubble is known to vary from $16Re^{-1}$ at low Reynolds number to $48Re^{-1}$ in the limit of very large Reynolds number \citep{Batchelor1967}. Hence, for nearly-spherical bubbles, the force balance implies $k(Re)Re=\frac{1}{12}Ga^2$, with $k(Re\ll1)=1$, $k(Re\gg1)=3$. The slow increase of $k(Re)$ in between these two limits, which follows approximately the power law $k(Re)\sim Re^{1/5}$ up to Reynolds numbers of a few hundreds, is the reason why the Reynolds number grows slightly more slowly than $Ga^2$ in the first regime. Bubbles are significantly distorted in the second regime which corresponds to the intermediate range $We=\mathcal{O}(1)$ ($We\approx0.6$ at $Re=200$ in water). Assuming that their frontal area, which is proportional to $\chi^{2/3}$, grows approximately linearly with the Weber number in that range yields $C_D(Re)\sim k(Re)Re^{-1}We\sim Re^{6/5}Mo^{1/3}Ga^{-2/3}$. The force balance then implies $Re\sim Ga^{5/6}Mo^{-5/48}$, a prediction supported by the numerical data for the various fluids in the $Ga$ range where path instability occurs.}}  
\begin{figure}
\vspace{5mm}
\centering
\subfloat[]{\includegraphics[width=0.25\textwidth,keepaspectratio]{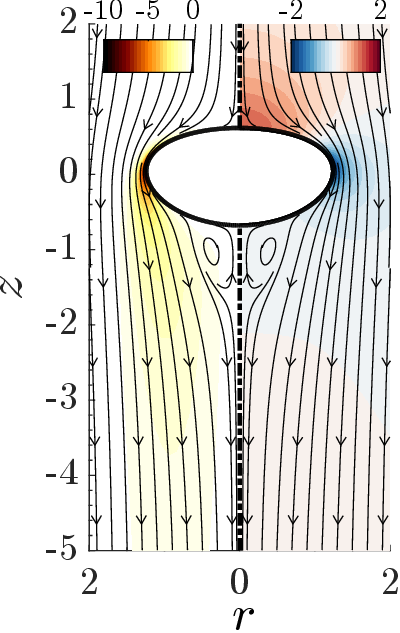}} \quad
\subfloat[]{\includegraphics[width=0.25\textwidth,keepaspectratio]{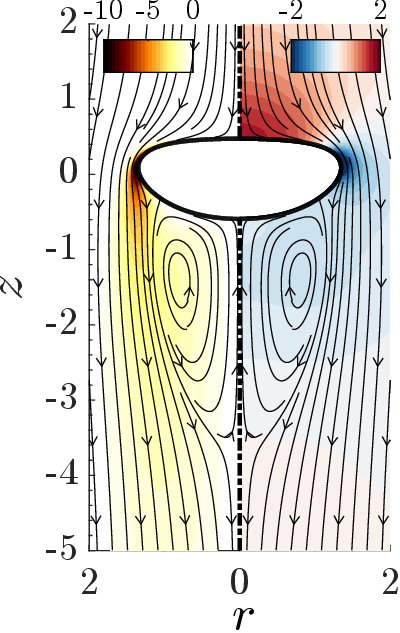}} \quad
\subfloat[]
{\includegraphics[width=0.25\textwidth,keepaspectratio]{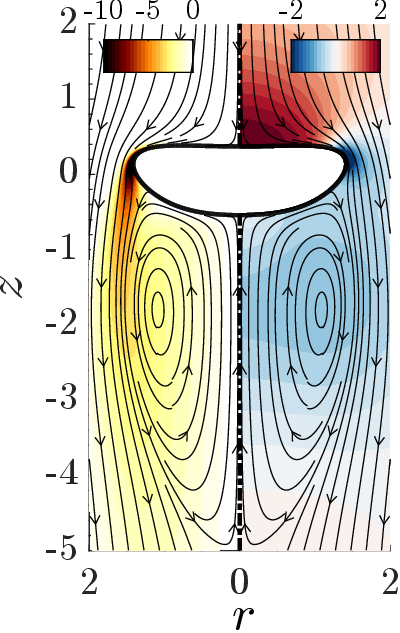}} \quad
\subfloat[]{\includegraphics[width=0.25\textwidth,keepaspectratio]{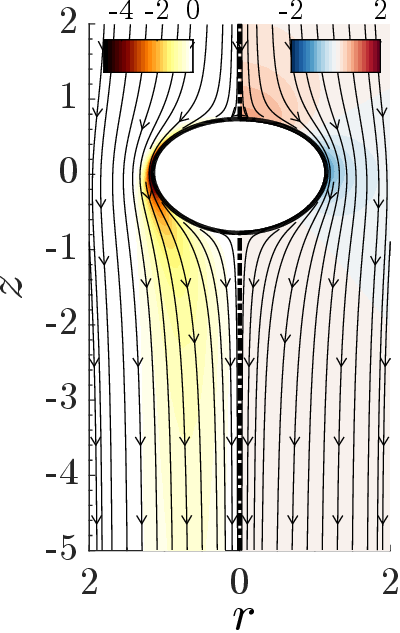}} \quad
\subfloat[]{\includegraphics[width=0.25\textwidth,keepaspectratio]{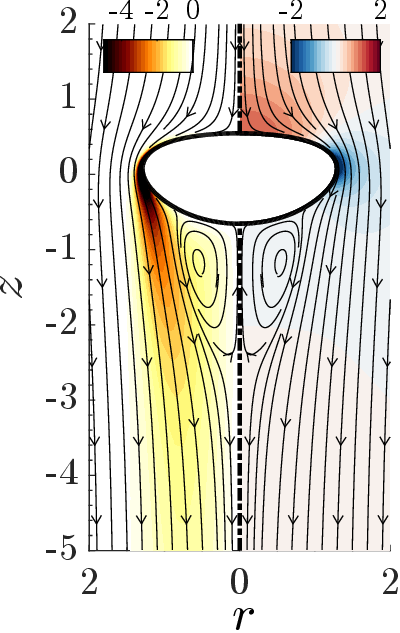}} \quad
\subfloat[]
{\includegraphics[width=0.25\textwidth,keepaspectratio]{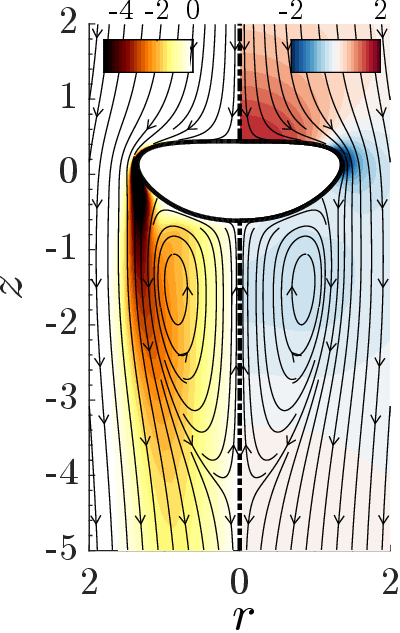}} \quad
\subfloat[]{\includegraphics[width=0.25\textwidth,keepaspectratio]{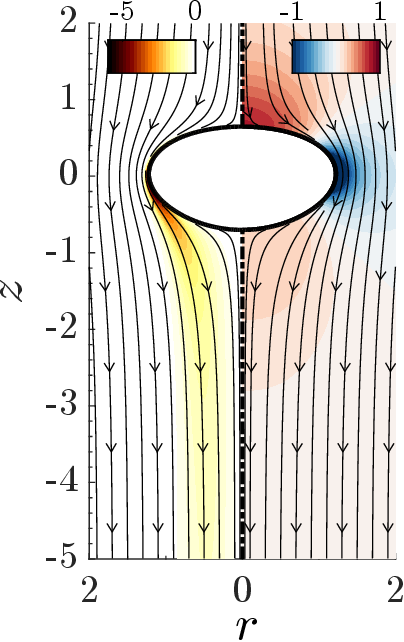}} \quad
\subfloat[]
{\includegraphics[width=0.25\textwidth,keepaspectratio]{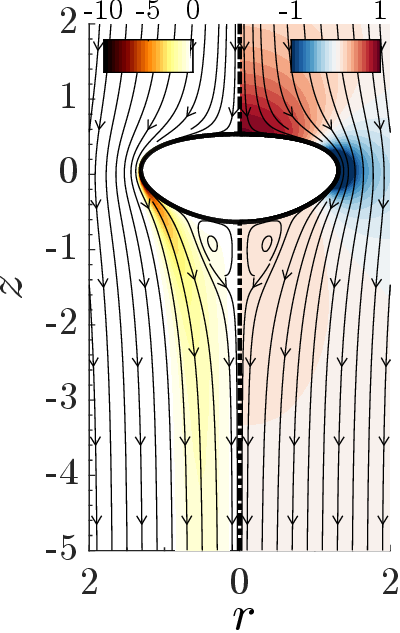}} \quad
\subfloat[] 
{\includegraphics[width=0.25\textwidth,keepaspectratio]{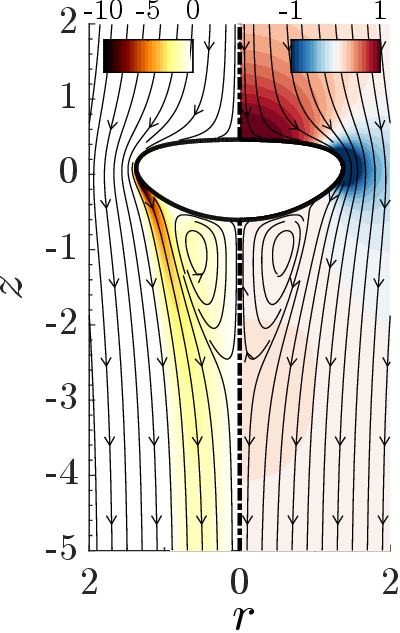}}
\caption{{\color{black}{Base flow past bubbles rising in three different liquids. $(a-c)$: DMST05 {\color{black}{($Mo\!=6.2\!\cdot 10^{-7}$)}}, with, from left to right, $Bo=3, 5$ and $7$; $(d-f)$: DMST02 {\color{black}{($Mo=1.6\times10^{-8}$)}}, with $Bo=1, 2$ and $3$; $(g-i)$: water at $20^\circ\,$C {\color{black}{($Mo\!=\!2.54\cdot 10^{-11}$)}}, with $Bo=0.4, 0.6$ and $0.8$. The left and right halves of each subfigure display the azimuthal vorticity and pressure distributions, respectively. The thin lines are the streamlines in the reference frame rising with the bubble.}} }
\label{fig:ContoursBaseFlow_DMST05}
\end{figure}
\subsection{Flow structure}
\label{structure}
\indent The flow structure in the base state reveals several interesting features. Figure \ref{fig:ContoursBaseFlow_DMST05} displays the azimuthal vorticity (left half of each panel) and pressure (right half) and distributions past a bubble rising in three different fluids in the $Bo$-range where the path instability threshold will later be shown to lie. With no surprise, the near-surface pressure distribution reaches its minimum very close to the equatorial plane, defined as the horizontal plane in which the longest axis of the bubble lies. This is a mere consequence of the increase of the fluid velocity along the interface from the apex of the bubble down to its equatorial plane. The norm of the azimuthal vorticity also reaches its maximum in that plane because this quantity is proportional to the product of the interface curvature in the vertical diametrical plane and the relative fluid velocity along the interface \citep{Batchelor1967}, both of which are maximum there. This vorticity results directly from the shear-free condition obeyed by the fluid at the interface and is responsible for the existence of a boundary layer that encapsulates the bubble and turns into a wake downstream of it \citep{Moore1963,Moore1965}. The wake structure is found to depend dramatically on the Bond number and, for a given $Bo$, on the fluid properties (compare subfigures $(a)$ and $(f)$, both for $Bo=3$). In water, all streamlines are open for $Bo=0.4$ (subfigure $(g)$) and they remain so up to $Bo\approx0.5$. Beyond this point, a standing eddy  exists and grows sharply with the Bond number, its length becoming of the same order as that of the long bubble axis for $Bo=0.8$ (subfigure $(i)$). Qualitatively similar trends are observed in the more viscous fluids.  However, for a given $Bo$, the normalized rise speed $U=u_b/(gD)^{1/2}$ is significantly smaller, and so is the Weber number $We\equiv\rho u_b^2D/\gamma=Bo\,U^2$. Indeed, keeping $Bo$ unchanged, the Weber numbers in two different fluids, 1 and 2, obey the relation $We_2/We_1=(Re_2/Re_1)^2(Mo_2/Mo_1)^{1/2}$. Hence, in the nearly five times more viscous silicone oil DMS-T05 for instance, the Weber number of a bubble with $Bo=0.5$ is eight times smaller than in water, leaving it nearly spherical. Consequently, compared with water, much larger $Bo$ are required for the tangential fluid velocity and interface curvature to be large enough for a standing eddy to develop at the back of the bubble. In the case of DMS-T05, the required conditions are achieved only for $Bo\gtrsim2.4$. \\
\begin{figure}
    \centering
\input{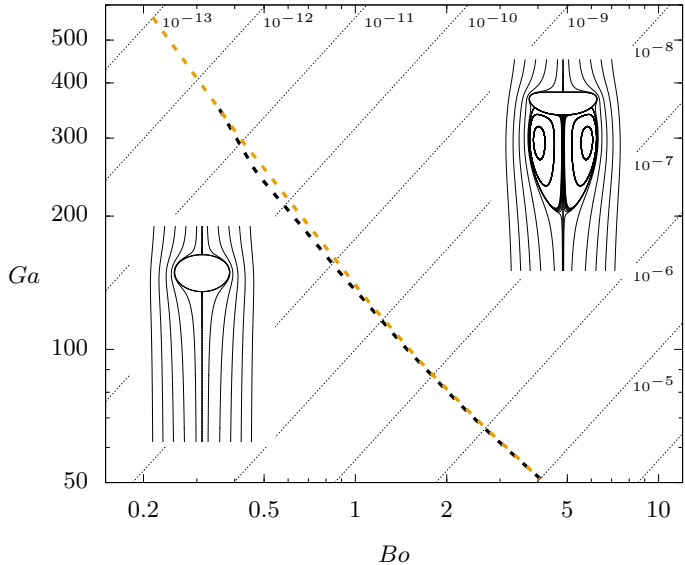}
  \vspace{-2mm}\\
\caption{Critical curve corresponding to the onset of a recirculating region at the back of the bubble in the $(Bo,Ga)$ plane. Yellow line: present results; black dashed line: predictions of \cite{Cano2013}. The left and right inserts show some streamlines around a bubble with $Bo=0.3$ in water and a bubble with $Bo=6.5$ in DMS-T05, respectively. The thin dotted lines correspond to constant values of the Morton number, i.e. to a given liquid; $Mo$ values are specified along the upper and right sides of the figure.}
 \label{fig:recirc}
\end{figure}
\indent The above information regarding the existence of a standing eddy in the various liquids is summarized in figure \ref{fig:recirc}. This plot shows the critical curve separating the upper region of the $Bo-Ga$ plane where a standing eddy exists, from the lower region in which no such flow structure takes place; the two insets display an example of each situation. Present predictions are seen to be in excellent agreement with those of the fully-resolved axisymmetric simulations carried out with \textit{Gerris} by \cite{Cano2013}. It will become clear in the next section that, in low-$Mo$ fluids, path instability arises for $(Ga,Bo)$ pairs located \text{below} this critical curve. This finding is of special importance in that it establishes that the existence of a standing eddy is not a prerequisite for the path of a freely rising bubble to become unstable, unlike the case of a fixed bubble \citep{Tchoufag2013}. More generally, it is well established that the initial `seed' leading to global wake instability past bluff bodies held fixed in a uniform stream is the growth of disturbances in the core of the standing eddy \citep{Chomaz2005}. That path instability may happen in the absence of such a near-wake structure proves that, in the relevant $Mo$-range, the mechanism that governs this instability is not driven by wake instability \textit{per se}. This confirms the findings of \cite{Cano2016} who established that, for $Mo\lesssim10^{-9}$, the path of bubbles with a frozen but realistic shape becomes unstable while their wake is still stable, the low-frequency `aerodynamic' mode being destabilised first.

\section{Neutral curves}
\label{neutral}
\subsection{Critical bubble size}
\label{size}
\begin{figure}
\centering
{\input{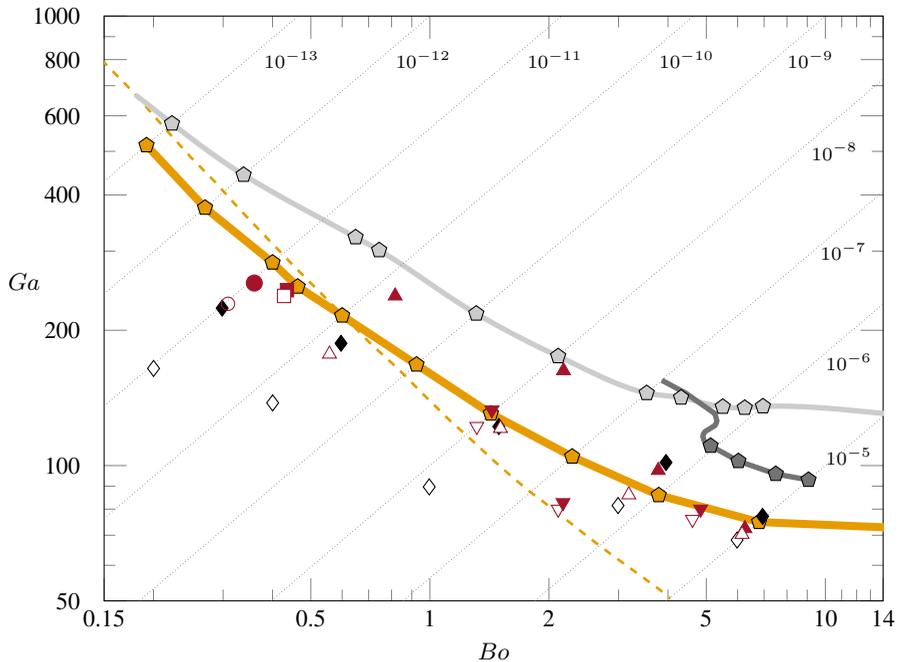}}
\caption{Neutral curve in the $(Bo,Ga)$ plane, obtained by connecting the $\textcolor{orange}{\pentagofill}$ symbols at which the threshold was determined. 
The pale and dark grey lines are the {\color{black}{two branches of the}} neutral curve obtained by considering the same base state but keeping the bubble shape frozen in the stability analysis; the thin yellow dashed line is the critical curve of  figure \ref{fig:recirc} beyond which a standing eddy exists at the back of the bubble in the base state. The thin dotted lines are iso-Morton number lines corresponding to a given liquid; $Mo$ values are specified along the upper and right sides of the figure. 
In a given fluid, open (resp. closed) symbols indicate stable (resp. unstable) paths observed in the simulations of \cite{Cano2016b} ($\rhombusdot$,$\rhombusfill$), in the experiments of \cite{DeVries2002}  ($\textcolor{purple}{\circletdot}$,$\textcolor{purple}{\circletfill}$) and \cite{Duineveld1995} ($\textcolor{purple}{\squaddot}$,$\textcolor{purple}{\squadfill}$) in ultrapure water (performed at temperatures of $28^\circ$C and $20^\circ$C, respectively), and in those of \cite{Zenit2008} ($\textcolor{purple}{\trianglepadot}$,$\textcolor{purple}{\trianglepafill}$) and \cite{Sato2009} ($\textcolor{purple}{\trianglepbdot}$,$\textcolor{purple}{\trianglepbfill}$) in various silicone oils.} 
\label{fig:neutralcurve}
\end{figure}
\begin{figure}
\vspace{5mm}
\centering
{\includegraphics[width=0.7\textwidth,keepaspectratio]{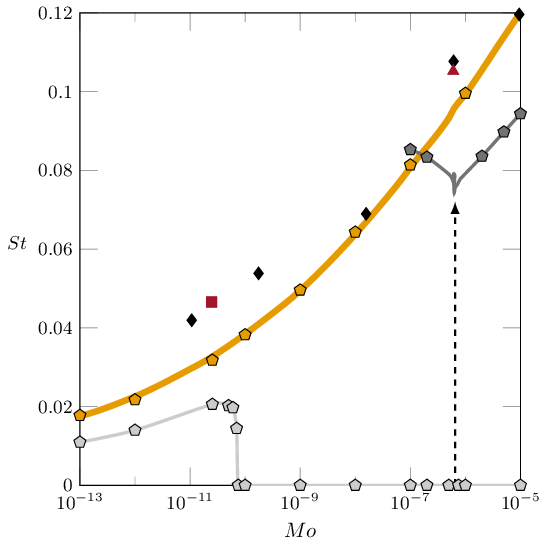}} 
     \caption{Variation of the reduced frequency $St=\lambda_i D/(2\pi u_b)$ at the threshold vs the Morton number. The yellow curve was obtained by connecting the $\textcolor{orange}{\pentagofill}$ symbols at which the frequency was determined. The pale and dark grey lines correspond to the predictions {\color{black}{along the two branches of the neutral curve}} predicted by the `frozen-shape' approximation (FSA) using the same base state (for readability, only five of the computed points are highlighted with a marker on the dark grey line). The vertical arrow at $Mo=2.7\times10^{-7}$ indicates the frequency jump associated with the switching from the stationary mode to the oscillatory mode predicted by FSA. 
     $\rhombusfill$: numerical predictions of  \cite{Cano2016b}; $\textcolor{purple}{\squadfill}$: experimental data of \cite{Duineveld1994} for $Bo=0.54$ in ultrapure water; $\textcolor{purple}{\trianglepafill}$: experimental data of \cite{Zenit2009} for $Bo=3.92$ in DMS-T05. }
 \label{fig:St}
\end{figure}

Using the procedure described in \S\,\ref{numer}, we selected a set of $Mo$ values, some of which correspond to specific liquids. For each of these values, we increased the bubble size, i.e. the Bond and Galilei numbers, until the real part of one of the eigenvalues associated with the non-axisymmetric modes $m=\pm1$ changes sign, which determines the threshold of path instability. The resulting neutral curve, obtained by linking these thresholds, is shown in figure \ref{fig:neutralcurve}. Together with figure \ref{fig:St}, this is presumably the most important result of the present study with respect to the original physical problem. For each fluid characterized by a Morton number in the range $10^{-13}-10^{-5}$, this curve readily provides the size of the smallest bubble whose vertical path becomes unstable. For instance, the path of a bubble rising in water at a temperature of $20^\circ$C becomes unstable at $Bo=0.463$ and $Ga=250$, yielding a critical diameter $D\approx1.854\,$mm \citep{Bonnefis2019,Bonnefis2023}. Similarly, the critical Bond numbers for Galinstan ($Mo=1.4\times10^{-13}$) and DMS-T11 ($Mo=9.9\times10^{-6}$) are {\color{black}{$0.19$}} and $6.85$, respectively. Therefore, the critical bubble sizes for these two fluids located close to the bounds of the $Mo$-interval considered here are {\color{black}{$1.48\,$mm}} and $3.875\,$mm, respectively.\\
\indent In figure \ref{fig:neutralcurve} we also reported some experimental data obtained under controlled conditions ensuring that the carrying fluid obeys a shear-free condition at the gas-liquid interface. The reference data of \cite{Duineveld1995} in ultrapure water at $20^\circ$C yield a critical bubble diameter close to $1.82\,$mm, which differs by less than $2\%$ from the present prediction. \cite{DeVries2002} also determined the onset of path instability in a tank of ultrapure water with an average temperature of $28^\circ$C ($Mo=1.13\times10^{-11}$). However, a $1.1^\circ$C\,m$^{-1}$ stabilizing temperature gradient was established to visualize the wake thanks to optical index gradients. Influence of this stratification on path instability is unknown and might be at the origin of the $8\%$ difference between the experimentally determined critical bubble size and the present prediction. Comparisons with data obtained in four different silicone oils \citep{Zenit2008} and in three other silicone oils in which photochromic dye was added to visualize the wake \citep{Sato2009} prove satisfactory in the sense that the neutral curve lies generally within the ($Bo,Ga$)-interval whose lower (resp. upper) bound corresponds to the largest (resp. smallest) bubble for which a stable (resp. unstable) path could be identified experimentally. The only two exceptions correspond to a probable outlier in Sato's data for $Mo=2.35\times10^{-7}$, and to DMS-T11 (the most viscous oil) in which the path of a bubble $4\%$ smaller than the critical size predicted here was found to be unstable. Numerical data obtained by \cite{Cano2016b} with \textit{Gerris} in the transition region (but not necessarily in the close vicinity of the threshold) are also reported. These data are seen to be in good agreement with present predictions for fluids with $Mo\gtrsim10^{-8}$. In contrast, as the authors anticipated in their conclusions, these simulations underestimate the critical bubble size in fluids having a lower Morton number. The lower $Mo$ the larger the underestimate, which makes the critical diameter in water under-predicted by $15-20\%$. Since these simulations made use of an adaptive grid with $128$ cells per bubble diameter on both sides of the interface, this significant underestimate helps appreciate how demanding fully-resolved simulations of bubbly flows in low-viscosity fluids are.\\
\indent In figure \ref{fig:neutralcurve} we also reported the critical curve of figure \ref{fig:recirc} corresponding to the onset of a recirculating region at the back of the bubble in the base flow (yellow dashed line). The yellow solid and dashed lines are seen to cross each other at $Mo\approx10^{-10}$. For lower $Mo$, the critical bubble size at which the path destabilizes first is such that no standing eddy exists in the base state. For the reasons discussed in the previous section, this leads to the conclusion that wake instability cannot be responsible for path instability in such low-$Mo$ fluids. 
 The pale and dark grey lines in the figure also provide interesting insight into the role of time-dependent bubble deformations. To obtain {\color{black}{the neutral curve resulting from these two}} lines, we considered the same base state as above, and imported the corresponding interface shape and rise speed in the stability code used by \cite{Tchoufag2014b} and \cite{Cano2016}. In this code, the interface shape is kept frozen, i.e. the bubble can only move as a rigid body. {\color{black}{A similar neutral curve was computed by \cite{Cano2016} (black line in their figure 7), using a base state obtained with \textit{Gerris}, instead of the present global Newton method. Their neutral curve is similar to that based on the two grey lines in figure \ref{fig:neutralcurve} but the critical $Ga$ they found at a given $Bo$ stands consistently below that determined here, with differences ranging from $5\%$ for $Bo\geq5$ to $15\%$ for $Bo=0.5$. Again, the reasons for these differences are to be found in the limited spatial resolution of the \textit{Gerris}-based computations in high-$Ga$ low-$Bo$ configurations.}}\\
 \indent Since the yellow and grey lines in figure \ref{fig:neutralcurve} are based on strictly identical base states, time-dependent deformations not accounted for in the stability analysis used to produce the second of them are entirely responsible for the differences observed in the fate of the imposed disturbances. As the comparison of the thresholds predicted by the two approaches in a given fluid evidence, deformations always lower the threshold, i.e. they promote path instability. However, for fluids with Morton numbers less than $\approx5\times10^{-7}$, the lower $Mo$ the weaker this influence. For instance, the `frozen-shape' approximation, hereinafter referred to as FSA, over-predicts the critical bubble size by nearly $23\%$ for $Mo=10^{-7}$, but this overestimate is reduced by a factor of three in the case of Galinstan. FSA actually predicts that the neutral curve involves two distinct modes intersecting at $Mo\approx2.8\times10^{-7}$. 
In the narrow range $4\times10^{-7}\lesssim Mo\lesssim6.5\times10^{-7}$, the mode that becomes most unstable beyond this intersection (dark grey line) exhibits a S shape implying a destabilisation-restabilisation scenario. In contrast, the neutral curve resulting from the stability analysis carried out with freely-deforming bubbles remains single-valued throughout the whole range of Morton number. For larger $Mo$, the FSA prediction based on this second mode is also quite good, with for instance a $16\%$ over-prediction of the critical bubble size in DMS-T11. \\
\indent At this point it is also worth mentioning that \cite{Cano2016} established that the neutral curve predicted by FSA and that corresponding to the onset of `pure' wake instability for a bubble with the same frozen shape held fixed in a uniform stream coincide in the range $10^{-9}\lesssim Mo\lesssim6\times10^{-7}$. Out of this range, {\color{black}{the threshold values of the Bond and Galilei numbers corresponding to the onset of wake instability in a given fluid are consistently larger than those at which FSA predicts the occurrence of path instability}}. For $Mo\lesssim10^{-9}$, the difference between the two thresholds increases gradually as $Mo$ is decreased. For $Mo\gtrsim6\times10^{-7}$, the threshold of wake instability coincides with the upper neutral branch captured by FSA (here the pale grey line). In other words, depending on the Morton number, the threshold corresponding to the onset of wake instability past a fixed bubble is either identical to or larger than that of path instability predicted by FSA. These findings, combined with the fact that figure \ref{fig:neutralcurve} reveals that FSA overestimates the threshold of path instability whatever $Mo$, establish that the wake of a fixed bubble with a frozen shape is always stable at the actual onset of path instability. Hence, path instability cannot be explained on the sole basis of the mechanism responsible for wake instability. \\
\subsection{Frequency at threshold}
\label{frequency}
The most spectacular qualitative difference between the predictions of the present approach and those of FSA is seen in figure \ref{fig:St}. This figure displays the Strouhal number, or reduced frequency, corresponding to the first unstable mode at the onset of the instability, computed as $St=\lambda_iD/(2\pi u_b)$, with $\lambda_i$ the imaginary part of the unstable eigenvalue at the threshold. As the yellow curve evidences, the present approach predicts that the most unstable mode of the system is always oscillatory, in line with experimental observations. The reduced frequency increases continuously with the Morton number, ranging from $St\approx0.02$ for $Mo=10^{-13}$ to $St\approx0.12$ for $Mo=10^{-5}$. FSA predicts that the first non-vertical path of the bubble is associated with `low-frequency' oscillations in low-$Mo$ liquids ($Mo\lesssim7\times10^{-11}$) and with `high-frequency' oscillations in liquids with $Mo\gtrsim3\times10^{-7}$, in agreement with the conclusions of  \cite{Cano2016}. Present findings with deformable bubbles are qualitatively consistent with these predictions in both ranges. The reduced frequencies determined in both approaches for low-$Mo$ liquids are even in fairly good quantitative agreement. This confirms that deformations do not bring major changes in the dynamics of isolated bubbles rising in such liquids, {\color{black}{which extends to all low-$Mo$ fluids the conclusions drawn by \cite{Bonnefis2023} in the specific case of water}}. More surprisingly, the agreement is still reasonable in fluids with $Mo\gtrsim3\times10^{-7}$, where the FSA neutral curve (now corresponding to the dark grey line) underestimates the reduced frequency by only roughly $20\%$. In contrast, the two approaches dramatically disagree in the intermediate $Mo$-range, where FSA predicts that the most unstable mode (pale grey line)  is stationary. Hence, one has to conclude that, in the linear framework adopted here, time-dependent deformations promote the instability of the oscillatory mode and succeed in making it more unstable than the stationary mode in this intermediate range, while only the latter mode may become unstable if these deformations are ignored. Some experimental and numerical data are also reported in figure \ref{fig:St}. Most of them were obtained significantly above the threshold, which explains why the corresponding frequencies lie above present predictions {\color{black}{(for instance, the bubble size is $8\%$ beyond the threshold value in the two experimental determinations)}}. Only the two numerical determinations of \cite{Cano2016b} in silicone oils DMS-T02 ($Mo=1.6\times10^{-8}$) and DMS-T11 ($Mo=9.9\times10^{-6}$) correspond to near-threshold conditions according to figure \ref{fig:neutralcurve}. The corresponding two $St$ values are seen to agree very well with present predictions.\\
\begin{figure}
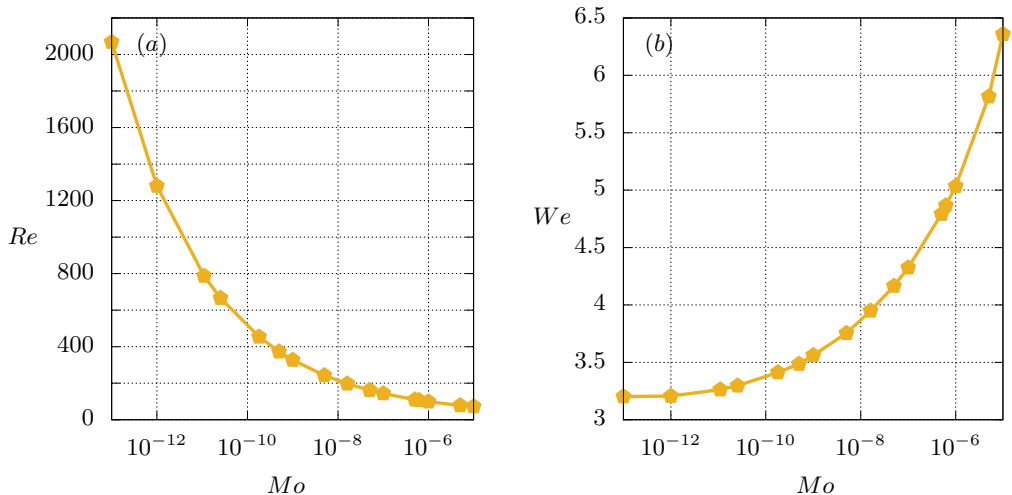

\centering
    {\input{Figures/Re-vs-Mo_Neutral-curve.tex}\input{Figures/We-vs-Mo_Neutral-curve.tex}}
 \vspace{-68mm}\\
\hspace{-26mm}$(a)$\hspace{63mm}$(b)$
\vspace{60mm} 
      \caption{Influence of the fluid properties on $(a)$: the critical Reynolds number, and $(b)$: the critical Weber number. }
\label{fig:ReWeMo} 
\end{figure}
\indent Having determined the normalized rise speed in the base state, we can re-plot the neutral curve of  figure \ref{fig:neutralcurve} in two different forms, to make the variations of the critical Reynolds number, $Re=Ga\,U$, and Weber number, $We=Bo\,U^2$, with the liquid properties apparent. 
Figure \ref{fig:ReWeMo}$(a)$ shows that the critical Reynolds number decreases sharply as $Mo$ increases, from $\approx2070$ for $Mo=10^{-13}$ to $70$ for $Mo=10^{-5}$, \textit{via} $Re=668$ for water at $20^\circ$C. Conversely, the critical Weber number (figure \ref{fig:neutralcurve}$(b)$) is seen to increase gradually from $3.20$ for $Mo=10^{-13}$ to $6.36$ for $Mo=10^{-5}$, \textit{via} $We=3.30$ for water. The above values for water are to be compared with those reported by \cite{Duineveld1995}, namely $Re=658$, $We=3.275$, from which they only differ by $1.5\%$ and $0.75\%$, respectively.
\section{Unstable modes: order of occurrence and physical nature}
\label{unstable}
In this section, we examine the variations of the most unstable eigenvalues of the coupled bubble-fluid system as the size of the bubble (here measured through the Bond number) is varied in the vicinity of the primary threshold. We also analyse the spatial structure of the associated global modes. It will soon become apparent that the order in which the unstable modes follow one another as the Bond number increases, and even in some cases the way the nature of a given mode varies, depend dramatically on the range of fluid properties under consideration, the complexity of the picture increasing as the Morton number is decreased. For this reason, we discuss these features in descending order of $Mo$.
\subsection{Silicone oil DMS-T05 {\color{black}{($Mo\!=6.2\!\times 10^{-7}$)}}}
\label{T05}
\subsubsection{First unstable eigenvalues: thresholds and nature of the associated modes} 
\label{eigen1}
\begin{figure}
\centering
\subfloat[]{\includegraphics[width=0.4\textwidth,keepaspectratio]{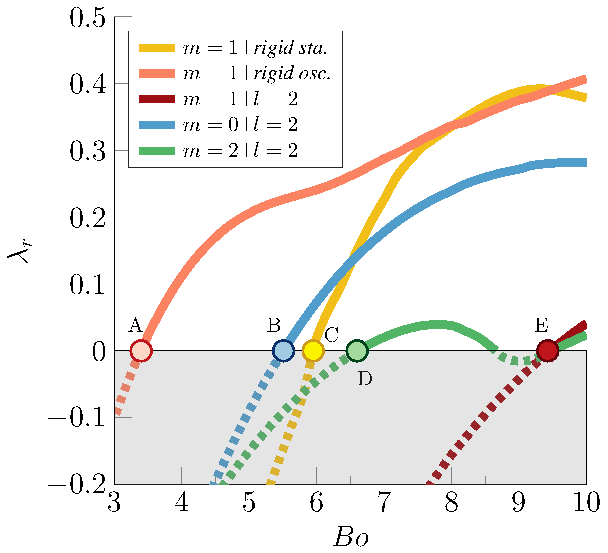}} \quad
\subfloat[]{\includegraphics[width=0.38\textwidth,keepaspectratio]{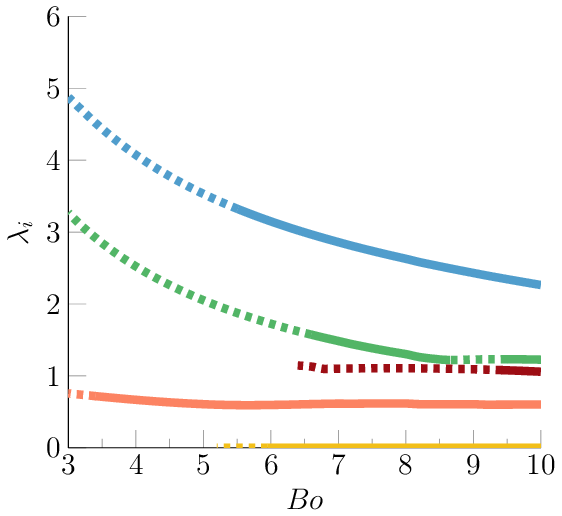}} 
\quad\subfloat[]{\includegraphics[width=0.14\textwidth,keepaspectratio]{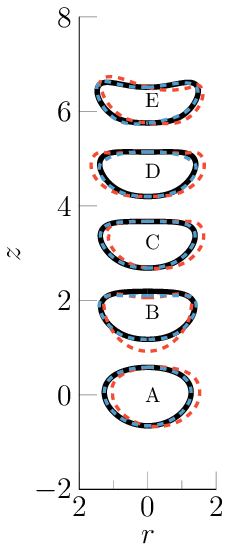}}
\caption{Variation of the first five unstable eigenvalues with the Bond number for bubbles rising in DMS-T05 {\color{black}{($Mo\!=6.2\!\times 10^{-7}$)}}. $(a)$: growth rate ($\lambda_r$), normalized by the gravitational time $(D/g)^{1/2}$; $(b)$: radian frequency ($\lambda_i$) normalized similarly; $(c)$: equilibrium shapes at the threshold (labels refer to the transition points marked with bullets in $(a)$). In $(a)-(b)$, dashed (resp. solid) lines represent the stable (resp. unstable) part of the various branches. In $(c)$, the bubble contour in the vertical diametrical plane is shown at the threshold of each of the five successive modes in the base state (black solid line) and, with an arbitrary amplitude of the disturbance, for $t=0$ (red dashed line) and $t=\pi/(2\lambda_i)$ (blue dashed line); bubble centroids are arbitrarily shifted in the $z$ direction for readability.}
\label{fig:LinearStabilityDMST05}
\end{figure}
Figure \ref{fig:LinearStabilityDMST05} displays the variations of the first five eigenvalues whose real part becomes positive as the the Bond number{\color{black}{, hence the size,}} of a bubble rising in silicone oil DMS-T05 is increased. All but one of these eigenvalues have a nonzero imaginary part, indicating that the corresponding instabilities arise through Hopf bifurcations. The threshold of path instability is encountered at $Bo=3.4$. This first unstable mode, associated with azimuthal wavenumbers $|m|=1$, has a dimensionless frequency $\lambda_i/2\pi\approx0.115$ at the threshold (throughout this section, eigenvalues are normalized using the gravitational time scale $(D/g)^{1/2}$). As subfigure $(c)$ suggests, this mode is essentially associated with a lateral drift of the bubble centroid accompanied by some rigid-body rotation. Time-dependent deformations are small: it will be shown in \S\,\ref{deform} that their maximum magnitude is barely $3\%$ of that of the horizontal displacement of the bubble centroid. For this reason, we refer to this mode as `rigid'. \\
\indent We continue to track this mode and other possible ones beyond this primary threshold by computing the base flow, still assumed axisymmetric, at the relevant Bond number. The growth rate of the rigid oscillatory mode continues to increase with $Bo$, and a second mode becomes unstable at $Bo\approx5.5$ (point B in figure \ref{fig:LinearStabilityDMST05}$(a)$). This mode is axisymmetric and oscillatory, with a dimensionless frequency approximately $5.5$ times higher than that of the rigid mode. Examination of the corresponding successive contours in subfigure $(c)$ reveals that this mode is associated with shape oscillations known as $l=2,m=0$, or $(2,0)$, in the terminology of spherical harmonics, i.e. the associated interface displacements leave the interface position unchanged at two angular positions located on both sides of the equatorial plane. We shall show in \S\,\ref{deform} that these deformations have a magnitude of the same order as the vertical displacement of the bubble centroid with respect to its reference position in the base state. For these reasons, we refer to this mode and those sharing similar properties as `shape' modes. By further increasing $Bo$, a third unstable mode is encountered at point C ($Bo\approx5.9$). Its growth rate increases strongly beyond the threshold so that this mode would quickly become dominant, were its amplitude similar to that of the primary oscillatory mode. Subfigure $(b)$ reveals that this mode is stationary and, since its azimuthal wavenumber is $|m|=1$, it is also asymmetric. These two properties imply that this mode induces an inclined path of the bubble centroid. Moreover, subfigure $(c)$ indicates that the corresponding interface deformations are weak compared with those associated with the previous `shape' mode, so that this stationary mode can be considered `rigid', just like the primary mode. 
Therefore, two rigid modes coexist in the system beyond $Bo\approx5.9$, but one is oscillatory while the other is stationary. To the best of our knowledge, steady oblique bubble paths have never been reported in experiments, suggesting that this second rigid mode never becomes dominant. Nonlinearities not taken into account here, i.e. the fact that the actual base flow at point C is no longer axisymmetric due to the nonlinear corrections brought by the primary oscillatory mode, and is presumably even not stationary owing to the shape oscillations associated with the $(2,0)$ shape mode, are likely the reason for this. \\
\indent A fourth mode becomes unstable at point D ($Bo\approx6.6$), and restabilizes itself within the narrow interval $8.65\lesssim Bo\lesssim9.45$ before becoming unstable again at larger $Bo$. This mode has much in common with the $(2,0)$ shape mode, except that it is not axisymmetric. Rather, it is associated with azimuthal wavenumbers $|m|=2$ and, in the $Bo$-range under consideration, its frequency is nearly half that of the $(2,0)$ mode. That the frequencies of two modes with the same $l$ but different $m$ differ in this context is no surprise, since spherical harmonics are no longer the eigenmodes with which a bubble oscillates when its undisturbed shape is not spherical \citep{Meiron1989}. Here the Bond number is large, and it is only in the small-$Bo$ limit that the two frequencies are expected to get close to each other. Due to the symmetries preserved by angular variations of the form $e^{\pm2i\theta}$, this $(2,2)$ mode does not induce any lateral displacement of the bubble. Last, at point E ($Bo\approx9.45$), a third shape mode with a frequency close to that of the $(2,2)$ mode becomes unstable. Deformations associated with this mode share the same characteristics as those induced by modes $(2,0)$ and $(2,2)$ on both sides of the bubble equatorial plane. In contrast, as subfigure $(c)$ makes clear, the two halves of the bubble deform in an asymmetric manner at every instant of time in the vertical diametrical plane, one half becoming more pointed while the other becomes more rounded. These characteristics identify this mode as the $(2,1)$ shape mode. Unlike those associated with $m=0$ and $|m|=2$, this mode may contribute to the lateral drift of the bubble at large enough Bond numbers, owing to its azimuthal asymmetry. It must be stressed that the full sequence described above is kept unchanged when the Morton number is increased up to $10^{-5}$. 
\\
\begin{figure}
\vspace{5mm}
\centering
{\includegraphics[width=0.49\textwidth,keepaspectratio]{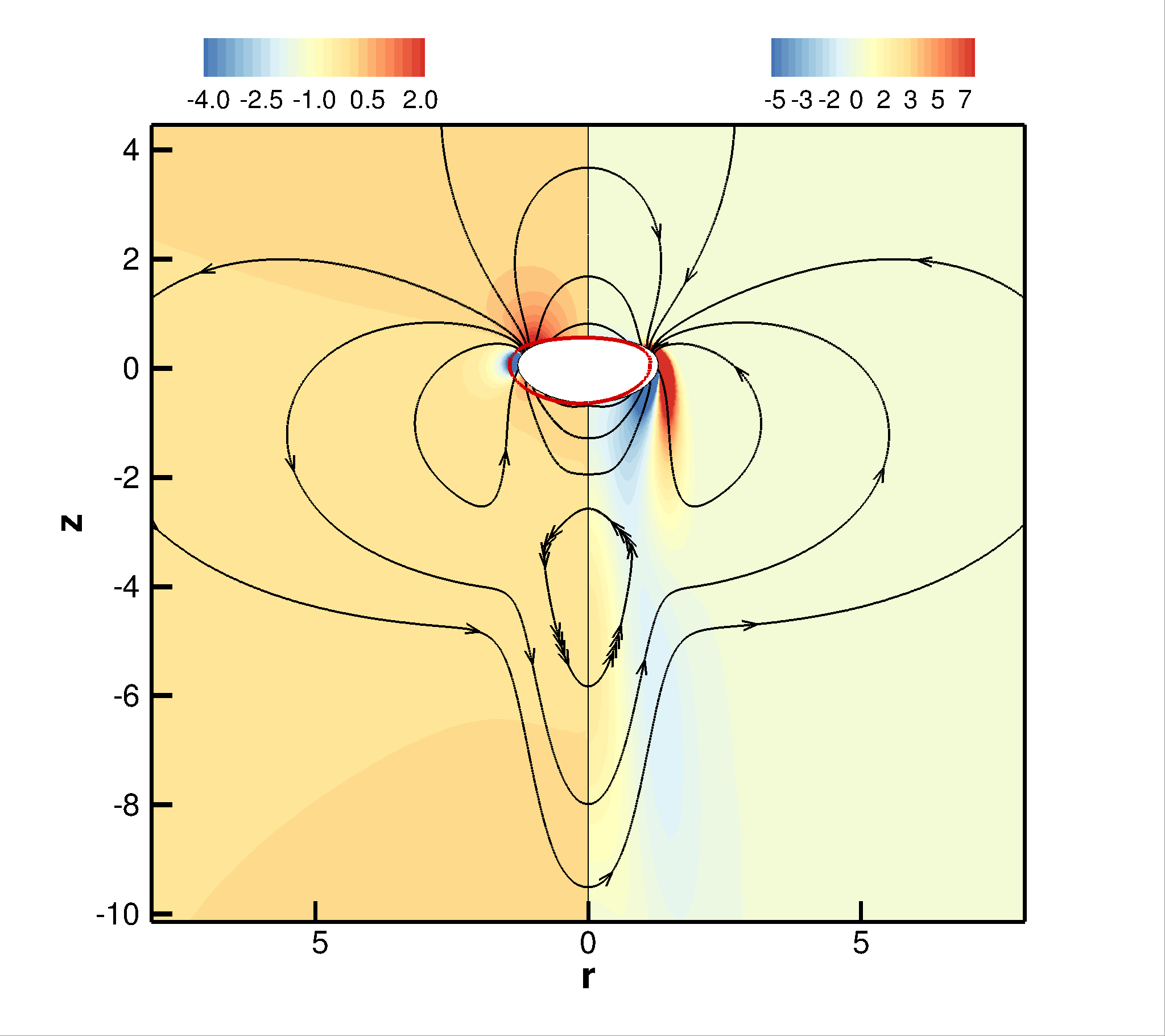}\includegraphics[width=0.49\textwidth,keepaspectratio]{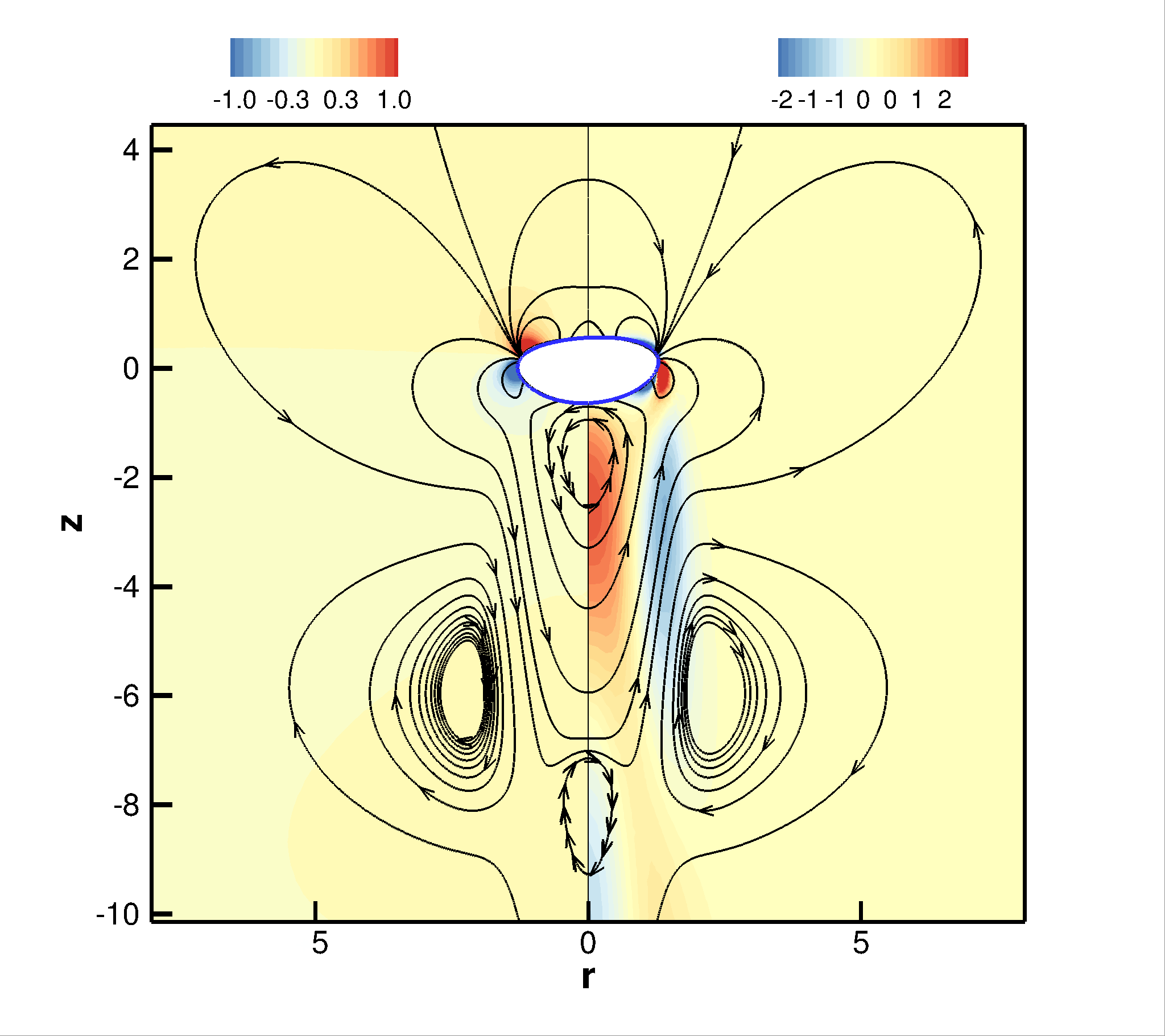}} \\
{\includegraphics[width=0.22\textwidth,keepaspectratio]{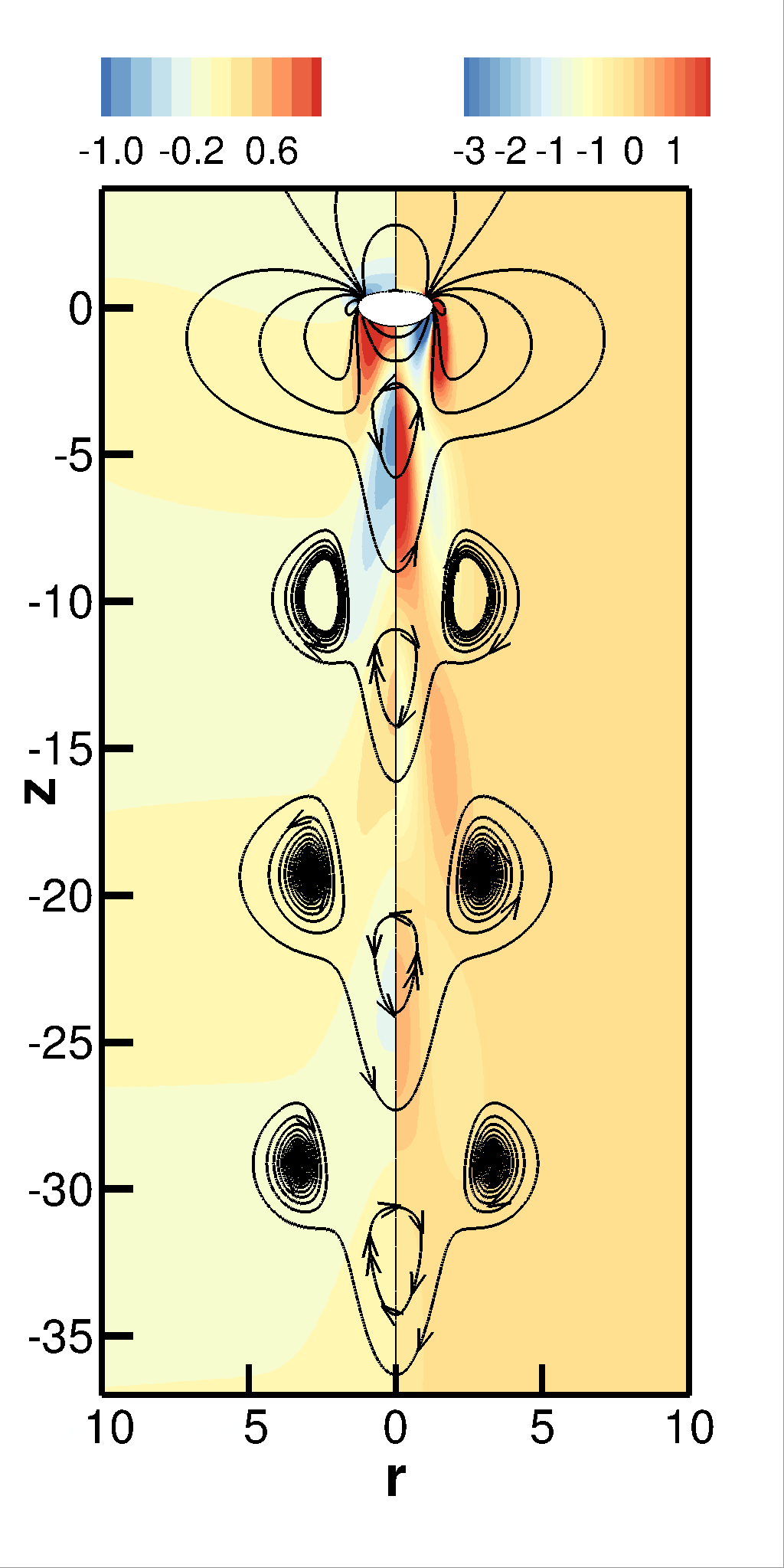}\quad\quad\includegraphics[width=0.22\textwidth,keepaspectratio]{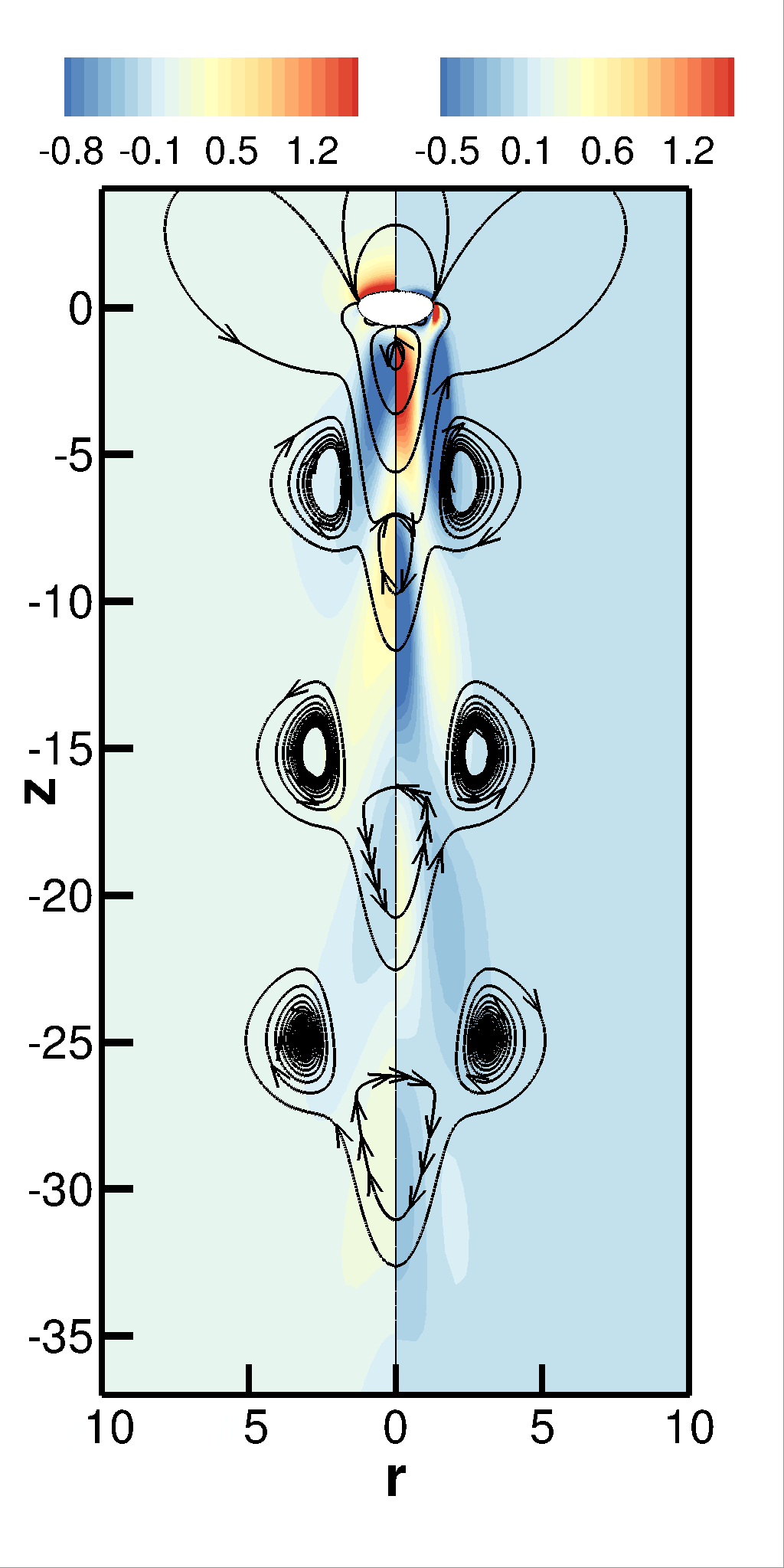}}
\vspace{-71mm}\\
\hspace{-41mm}$(a)$\hspace{61mm}$(b)$\\
\vspace{56mm} 
\hspace{-16mm}$(c)$\hspace{32mm}$(d)$\\
\vspace{7mm}
     \caption{Structure of the first unstable mode past a bubble rising in DMS-T05 at $Bo=3.7$. $(a)$ and $(c)$: real part of the mode near the bubble and further downstream in the wake, respectively; $(b)$ and $(d)$: same for the imaginary part. The left and right halves in $(a-b)$ display the pressure and azimuthal vorticity iso-levels, respectively; those in $(c-d)$ display the azimuthal velocity and vorticity iso-levels, respectively; some streamlines defined in the reference frame of the base configuration are also shown. Black, red and blue bubble contours correspond to the base state, and the real and imaginary parts of the interface disturbance ($\hat\eta$), respectively.
      }
 \label{fig:modesT05}
\end{figure}
\subsubsection{Spatial structure of the first unstable mode}
\label{spatial1}
Figure \ref{fig:modesT05} shows the spatial structure of the first unstable mode slightly above the threshold. The real and imaginary parts of a given mode may be thought of as the associated disturbance at two different instants of time a quarter of a period apart, i.e. shifted by an interval $\tau=\frac{\pi}{2}\lambda_i^{-1}$. Therefore, if the bubble is zigzagging in a vertical plane (as it does if the two disturbances associated with modes $m=+1$ and $m=-1$ have the same amplitude), the real part corresponds to the time instant by which the bubble reaches its maximal lateral excursion (red contours in figures \ref{fig:modesT05}$(a)$ and \ref{fig:LinearStabilityDMST05}$(c)$), while the imaginary part corresponds to that by which it crosses the midline of its path (blue contours in figures \ref{fig:modesT05}$(b)$ and \ref{fig:LinearStabilityDMST05}$(c)$). The streamlines pattern evidences the antisymmetric nature of the disturbance flow. An alternation of positive and negative vorticity zones may be noticed in the wake, both  in every horizontal plane (varying $r$) and at every radial position from the axis of the base flow (varying $z$). This pattern is indicative of a vortex shedding process. A specific structure in which fluid particles rotate clockwise on both sides of the axis of the base flow is visible near the bottom of subfigure $(b)$; in a three-dimensional representation, this structure would look like a pair of crescents connected through their horns in the vertical plane $\theta=\pm\pi/2$, with fluid particles rotating in the same direction in every azimuthal plane. As subfigures $(c-d)$ make clear, this structure is actually the first of a series that develop further downstream in the wake, with alternating positive and negative directions of rotation. 
\subsection{Silicone oil DMS-T02 {\color{black}{($Mo=1.6\times10^{-8}$)}}}
\begin{figure}
\vspace{5mm}
\centering
\subfloat[]{\includegraphics[width=0.4\textwidth,keepaspectratio]{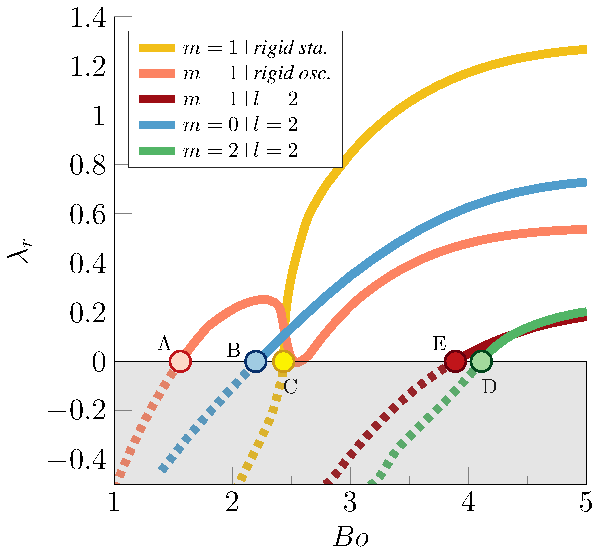}} \quad
\subfloat[]{\includegraphics[width=0.38\textwidth,keepaspectratio]{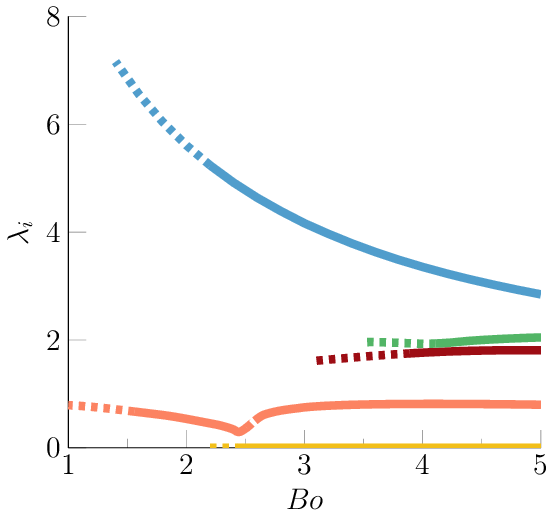}}
\quad\subfloat[]{\includegraphics[width=0.14\textwidth,keepaspectratio]{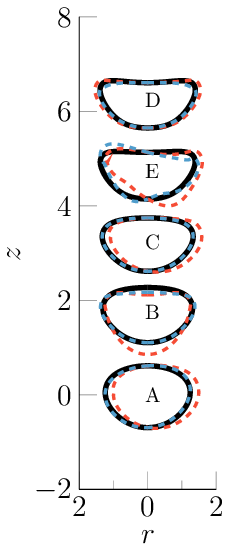}}
\caption{Same as figure \ref{fig:LinearStabilityDMST05} for bubbles rising in DMS-T02  {\color{black}{($Mo=1.6\times10^{-8}$)}}.}
\label{fig:LinearStabilityDMST02}
\end{figure}
We now move to silicone oil DMS-T02 which is $2.6$ times less viscous than DMS-T05 and has almost the same surface tension. Figure \ref{fig:LinearStabilityDMST02} shows how the first five eigenvalues yielding unstable modes vary with the Bond number in this case. The beginning of the sequence is similar to that described for DMS-T05. That is, path instability first arises through an oscillatory `rigid' mode that becomes unstable at $Bo=1.56$ (point A in subfigure $(a)$). Then the shape mode $(2,0)$ becomes unstable at point B ($Bo=2.2$), followed by the stationary rigid mode at point C  ($Bo=2.43$). Further increasing $Bo$, the $(2,1)$ and $(2,2)$ shape modes become unstable at points E ($Bo=3.89$) and D ($Bo=4.12$), respectively. Note that, compared with the case of DMS-T05, these last two modes emerge in reverse order. \\
\indent Nevertheless, the main originality of the present bifurcation diagram compared with that of DMS-T05 is that the growth rate of the primary mode starts to decrease slightly beyond point B, and becomes even very weakly negative within a tiny interval around $Bo\approx2.55$, before turning positive again and growing continuously up to the upper bound of the explored Bond number range. 
This non-monotonic behaviour is to be related to the fact that DMS-T02 stands in the middle of the $Mo$-range where FSA predicts the emergence of path instability through a stationary bifurcation instead of a Hopf bifurcation, being unable to detect that the oscillatory rigid mode becomes actually unstable first. This failure is presumably closely connected with the fact that, although interface deformations succeed in making the oscillatory mode unstable first, the growth rate of this mode remains quite low throughout the interval bounded by points A and C. We shall come back to this point in \S\,\ref{conclu}. 
\subsection{Water at $20^\circ$C {\color{black}{($Mo\!=\!2.54\times 10^{-11}$)}}}
\label{water}
\subsubsection{First unstable eigenvalues: thresholds and nature of the associated modes} 
\label{eigen2}
\begin{figure}
\subfloat[]{\includegraphics[width=0.4\textwidth,keepaspectratio]{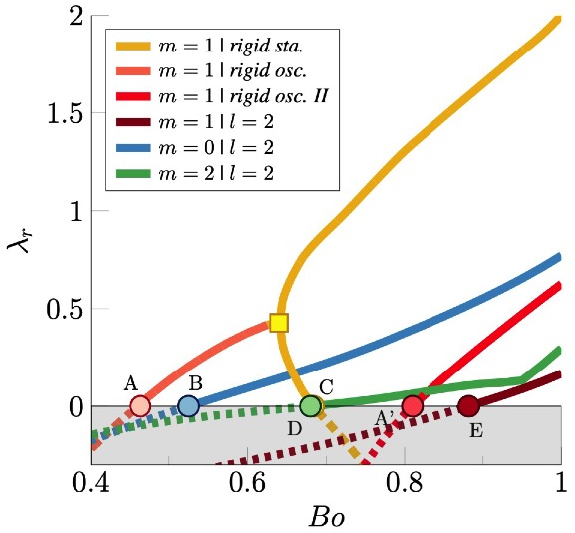}} \quad
\subfloat[]{\includegraphics[width=0.38\textwidth,keepaspectratio]{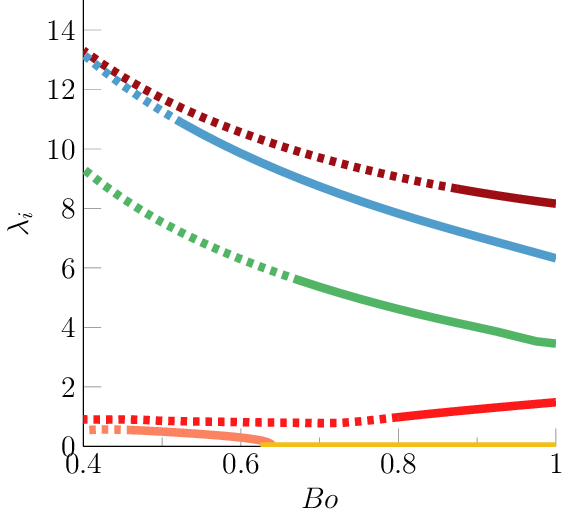}} \quad
\subfloat[]
{\includegraphics[width=0.14\textwidth,keepaspectratio]{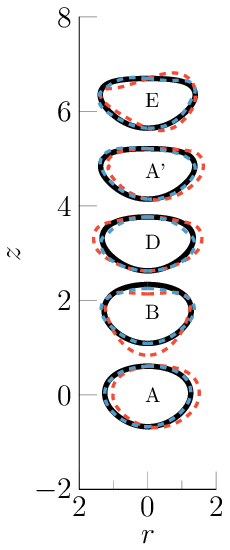}}
\caption{Same as figure \ref{fig:LinearStabilityDMST05} for bubbles rising in water at $20^\circ\,$C {\color{black}{($Mo\!=\!2.54\times 10^{-11}$)}}. In $(a)$, the square indicates the codimension-two `exceptional' point at which the pair of complex conjugate eigenvalues associated with the first unstable mode turns into a pair of real eigenvalues. }
\label{fig:LinearStabilityWater}
\vspace{-3mm}
\end{figure}

We finally consider the case of bubbles rising in pure water maintained at a temperature of $20^\circ\,$C. Variations of the eigenvalues leading to the first five unstable modes are shown in figure \ref{fig:LinearStabilityWater}. Again, the sequence begins with the emergence of an oscillatory rigid mode that becomes unstable at $Bo=0.463$ with a dimensionless frequency $\lambda_i/2\pi=0.088$, followed by the $(2,0)$ shape mode at $Bo=0.525$. However, the next steps differ deeply from those encountered in the previous two cases. The primary mode grows continuously until $Bo\approx0.64$. There, this mode splits into two separate branches, both of which are stationary. Therefore, at the point (marked with a square in the figure) where the three branches meet, the pair of complex eigenvalues associated with the oscillatory mode turns into a pair of real eigenvalues. Such a codimension 2 singular point is referred to as `exceptional' in the theory of non-Hermitian Hamiltonian systems \citep{Bergholtz2021}. {\color{black}{In the present case, this exceptional point, at which the two eigenvalues change their nature, corresponds to a Takens-Bogdanov bifurcation with $\mathcal{O}(2)$-symmetry \citep{Dangelmayr1987}.}} Beyond this point, the upper stationary branch exhibits a rapid and continuous growth. Conversely, the growth rate of the lower branch decreases sharply with the distance to the exceptional point, until it becomes negative at $Bo=0.685$ (point C in the figure). 
Similar to the cases of DMS-T02 and DMS-T05, and undoubtedly for the same reasons, inclined paths corresponding to the above stationary modes have not been observed in experiments carried out in water, irrespective of the bubble size. 
\\
\indent Next, figure \ref{fig:LinearStabilityWater} indicates that the mode $(2,2)$ becomes unstable at $Bo=0.68$ (point D in the figure) and then grows slowly with the Bond number. Then, a fourth mode becomes unstable at $Bo=0.809$ (point A') with a dimensionless frequency $\lambda_i/2\pi= 0.162$. This mode (identified as \textit{rigid osc. II} in the legend of figure \ref{fig:LinearStabilityWater}$(a)$) is associated with an azimuthal wavenumber $|m|=1$, exhibits low-frequency oscillations (subfigure $(b)$), and is not accompanied by significant shape oscillations (subfigure $(c)$). These characteristics point to another rigid mode. Hence, similar to the right part of figures \ref{fig:LinearStabilityDMST05} and \ref{fig:LinearStabilityDMST02}, two rigid modes coexist in the case of water beyond point A'. In all three cases, the most amplified of them is stationary (yellow curve in figure \ref{fig:LinearStabilityWater}) while the other (red curve) exhibits slow oscillations. Finally, figure \ref{fig:LinearStabilityWater} indicates that a fifth eigenvalue crosses the real axis at $Bo\approx0.88$ (point E). As subfigures $(b-c)$ make clear, this eigenvalue corresponds to the asymmetric shape mode $(2,1)$ already encountered with DMS-T02 and DMS-T05. One may notice that the mode $(2,1)$ becomes unstable at a larger $Bo$ than the mode $(2,2)$ in the cases of water and DMS-T05, while the reverse is true for DMS-T02. It is also worth noting that, among the three shape modes revealed by figure \ref{fig:LinearStabilityWater}, the mode $(2,2)$ is the one with the lowest frequency in the range of Bond numbers relevant here, say $Bo\gtrsim0.4$, followed by the mode $(2,0)$ and then the mode $(2,1)$. This is in line with the conclusions obtained by \cite{Meiron1989} assuming an inviscid potential flow. In contrast, figures \ref{fig:LinearStabilityDMST05} and \ref{fig:LinearStabilityDMST02} show that modes $(2,1)$ and $(2,2)$ have nearly equal frequencies, both significantly lower that that of the mode $(2,0)$, in DMS-T05 and DMS-T02. This suggests that vortical effects not accounted for in the above reference significantly modify the dynamics of shape oscillations of strongly deformed bubbles, even in low-viscosity fluids. Indeed, although the viscosity of DMS-T02 is less than twice that of water, its surface tension is four times lower, so that the relevant Bond numbers in DMS-T02 are typically larger than $3.0$, while those in water are less than $1.0$.\\
\indent To finish with figure \ref{fig:LinearStabilityWater}, it is important to stress that the succession of modes it reveals is the canonical sequence encountered throughout the low-$Mo$-range, say for $Mo\lesssim10^{-9}$. For instance, the same first five modes become unstable in the same order in the case of DMS-T00 ($Mo=1.8\times10^{-10}$), with the low-frequency rigid mode splitting into two stationary modes at an exceptional point, just as in figure \ref{fig:LinearStabilityWater}. Only the thresholds and frequencies differ, with points A-B then located at $Bo=0.66$ and $0.805$, followed by points C and D almost coincident at $Bo=0.98$, and finally points A' and E located at $Bo=1.135$ and $1.42$, respectively. 
\begin{figure}
\vspace{5mm}
\centering
{\includegraphics[width=0.49\textwidth,keepaspectratio]{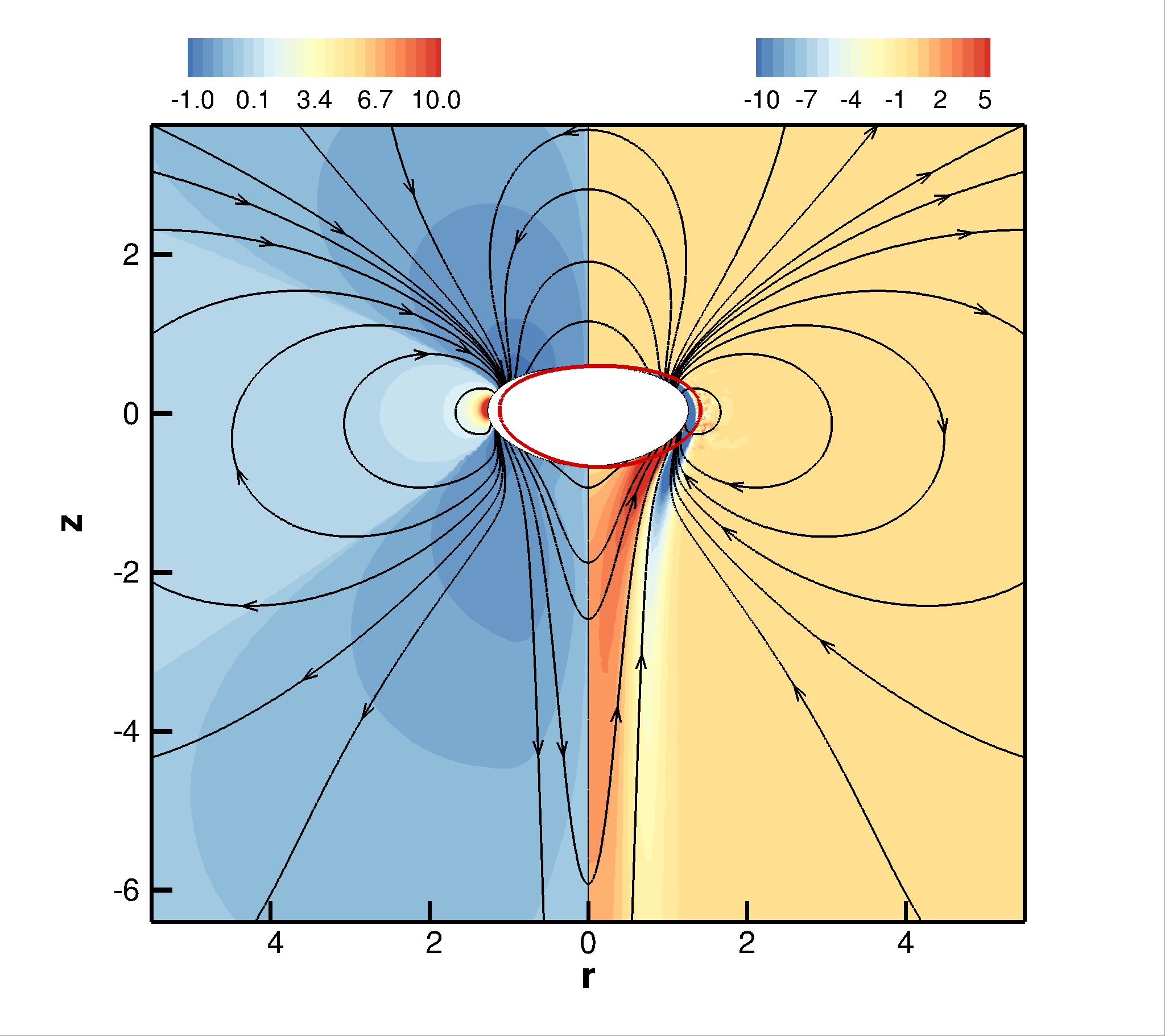}\includegraphics[width=0.49\textwidth,keepaspectratio]{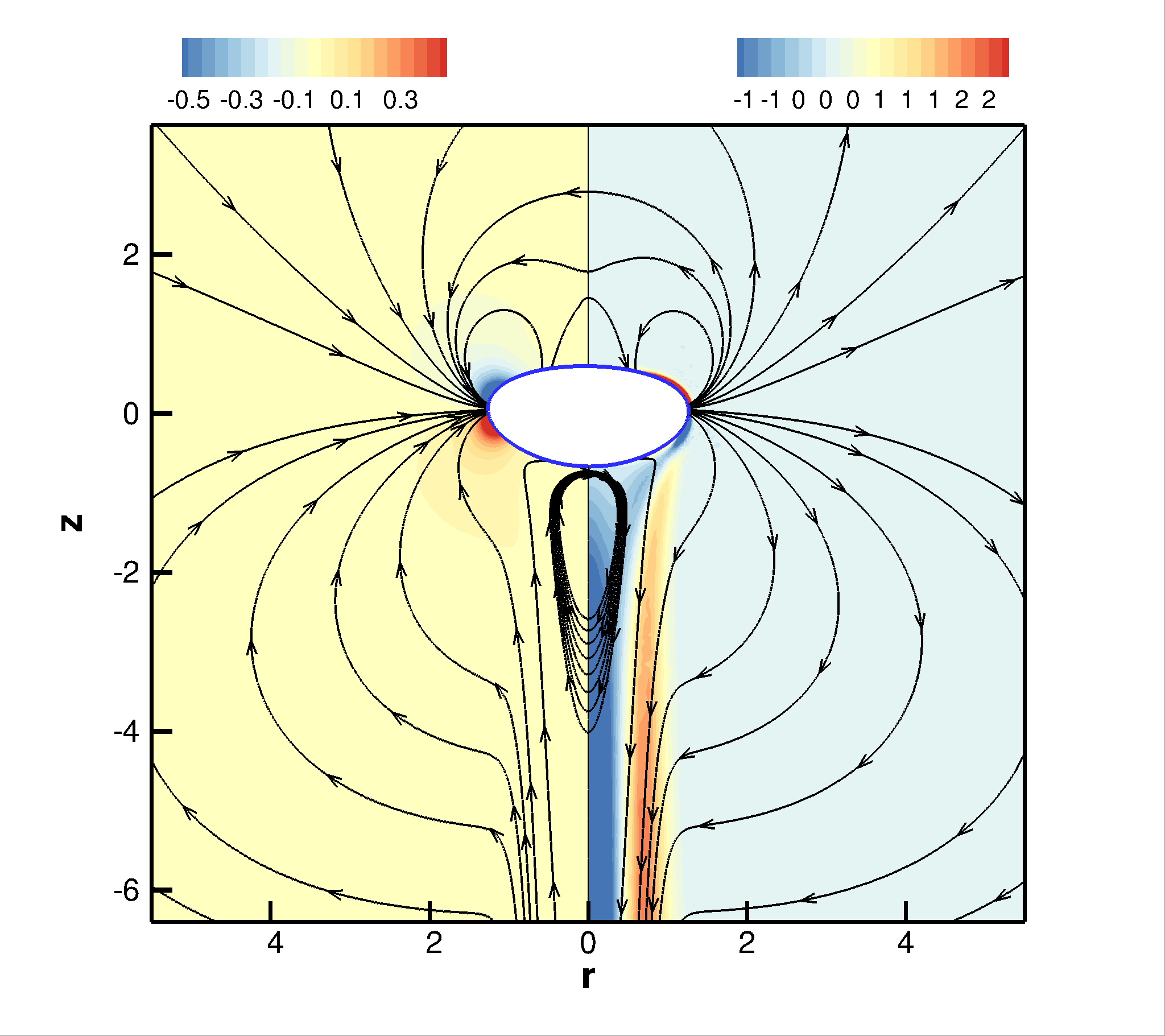}} \\
{\includegraphics[width=0.22\textwidth,keepaspectratio]{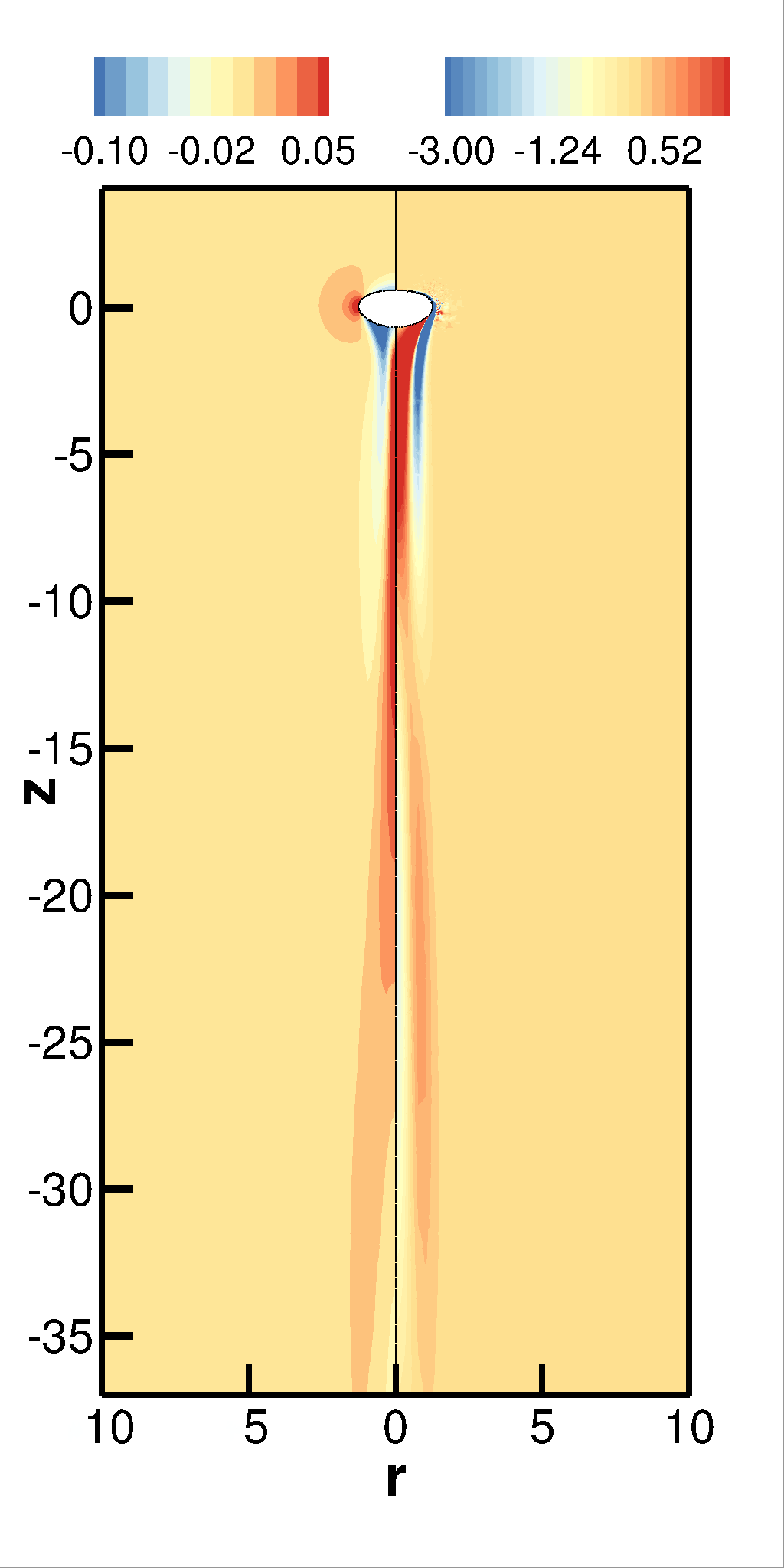}\quad\includegraphics[width=0.22\textwidth,keepaspectratio]{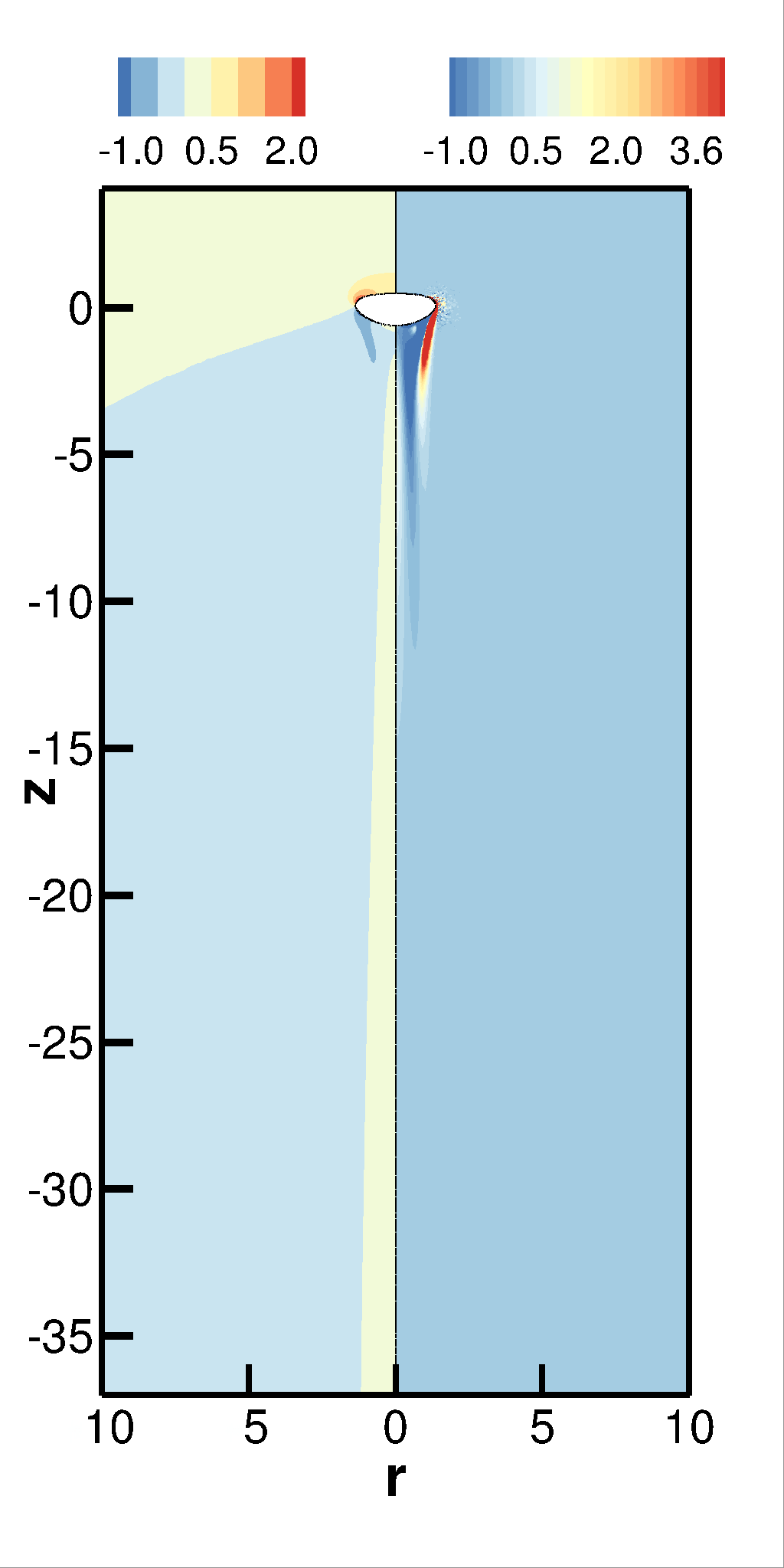}\quad\includegraphics[width=0.22\textwidth,keepaspectratio]{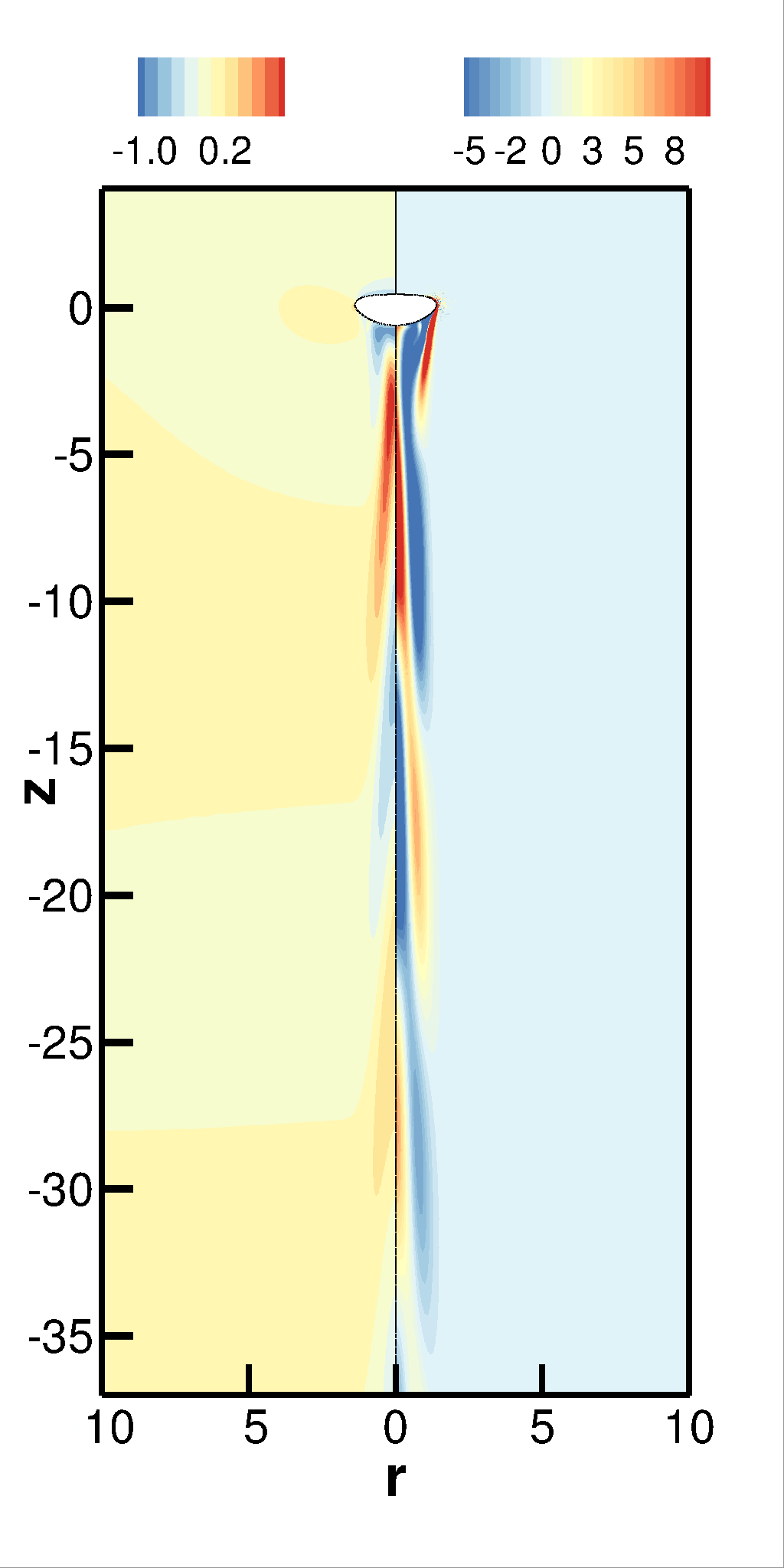}}
\vspace{-71mm}\\
\hspace{-41mm}$(a)$\hspace{61mm}$(b)$\\
\vspace{56mm} 
\hspace{-16.5mm}$(c)$\hspace{29mm}$(d)$\hspace{29mm}$(e)$\\
\vspace{6mm}
     \caption{Structure of the rigid modes past a bubble rising in water. Top: primary low-frequency mode at $Bo=0.48$ in the bubble vicinity, with the real and imaginary parts shown in $(a)$ and $(b)$, respectively. The left and right halves of the two subfigures display the pressure and azimuthal vorticity iso-levels, respectively; some streamlines defined in the reference frame of the base configuration are also shown. Black, red and blue bubble contours correspond to the base state, and the real and imaginary parts of the interface disturbance, respectively. Bottom: wake structure of the successive rigid modes. $(c)$: primary low-frequency mode at $Bo=0.48$; $(d)$: most amplified stationary mode at $Bo=0.75$; $(e)$: secondary oscillating mode at $Bo=0.85$. The right half of each subfigure displays the real part of the azimuthal vorticity iso-levels; the left half of $(c)$ and $(e)$ (resp. $(d)$) displays the real (resp. imaginary) part of the azimuthal velocity iso-levels.} 
 \label{fig:rigidmodesWater}
 \vspace{-3mm}
\end{figure}
\begin{figure}
\vspace{5mm}
\centering
{\includegraphics[width=0.49\textwidth,keepaspectratio]{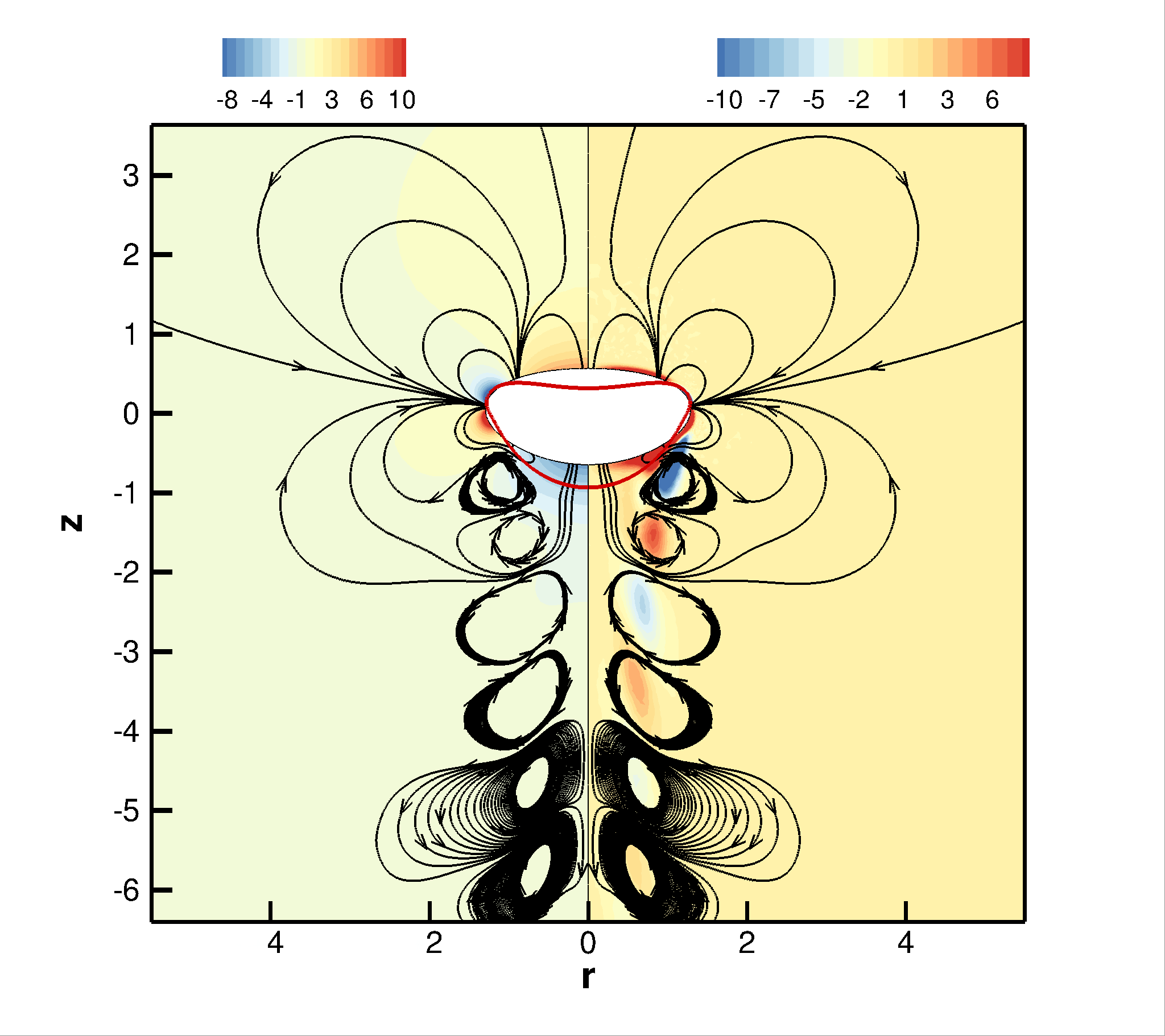}\includegraphics[width=0.49\textwidth,keepaspectratio]{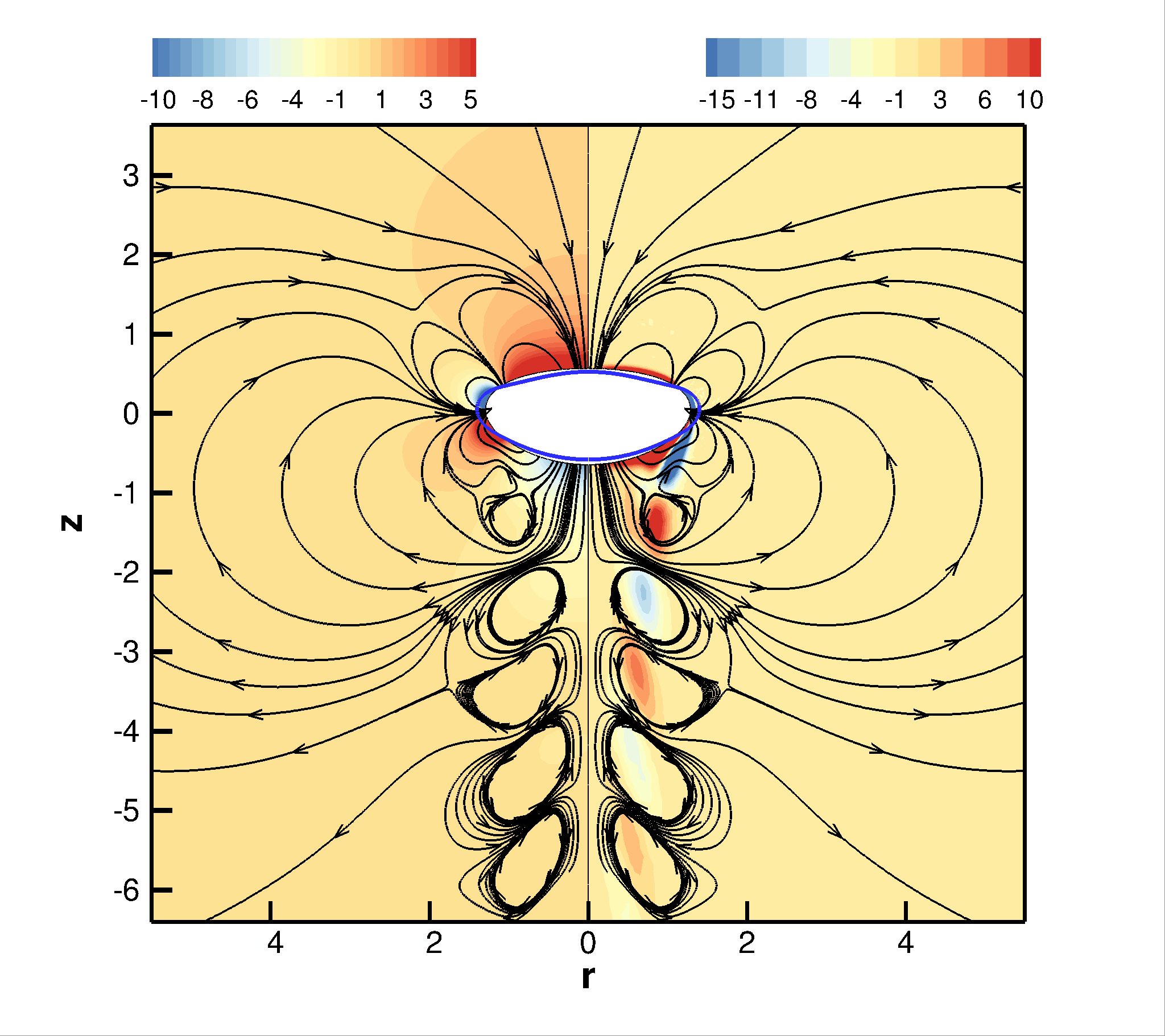}}\\
{\includegraphics[width=0.49\textwidth,keepaspectratio]{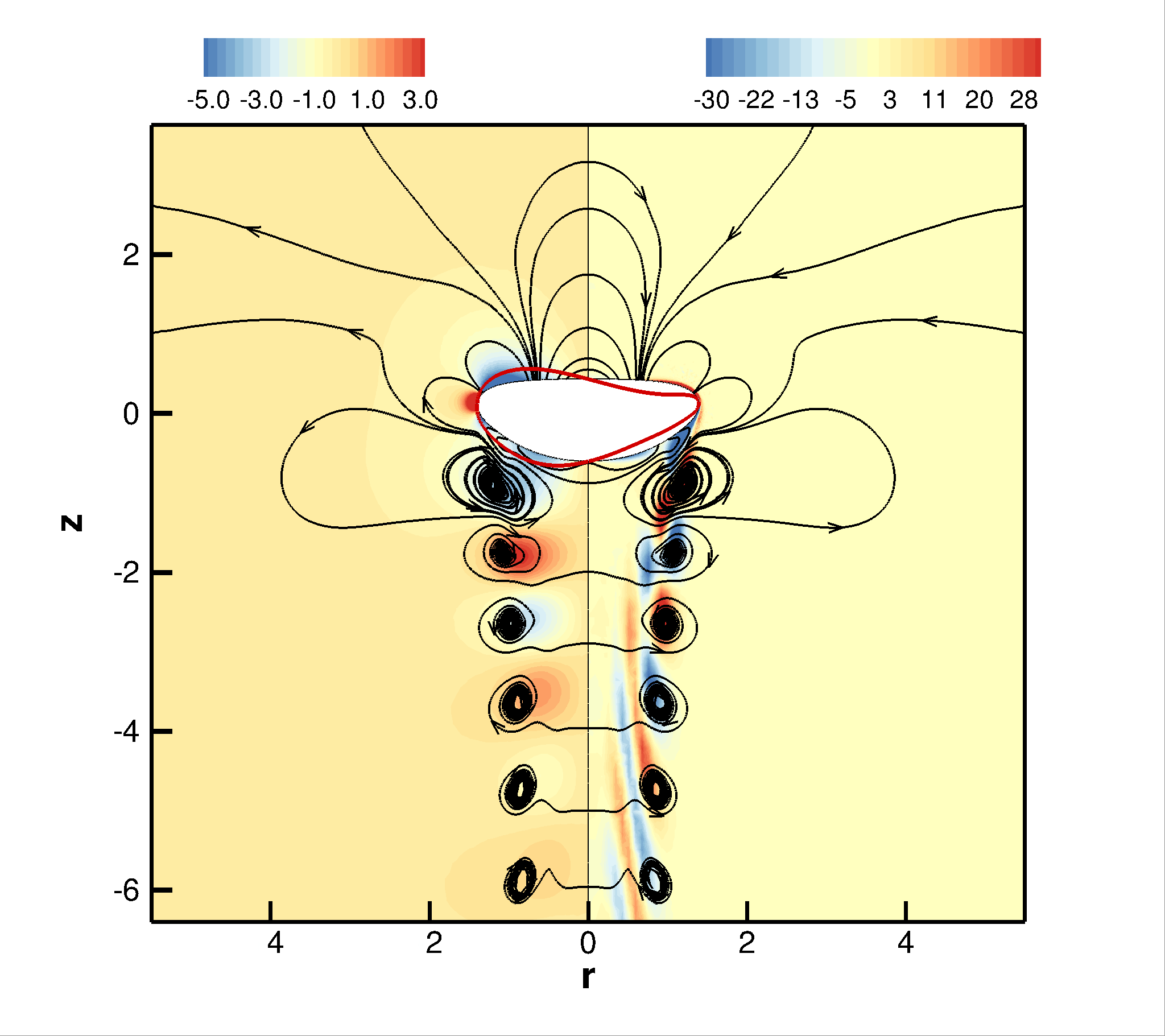}\includegraphics[width=0.49\textwidth,keepaspectratio]{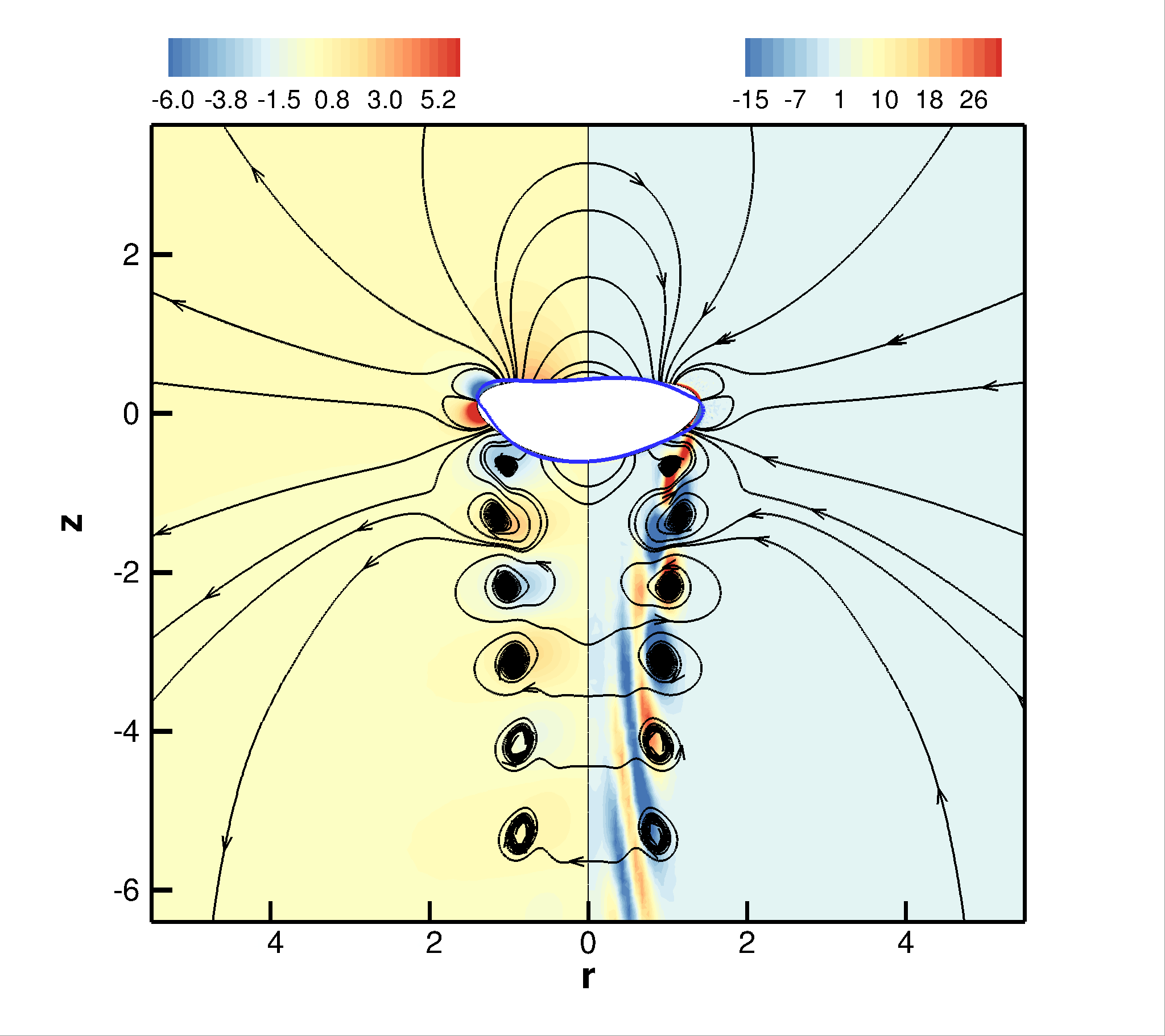}}
\vspace{-71mm}\\
\hspace{-41mm}$(a)$\hspace{61mm}$(b)$\\
\vspace{55.5mm} 
\hspace{-41mm}$(c)$\hspace{62mm}$(d)$\\
\vspace{7mm}
     \caption{Structure of the first axisymmetric and asymmetric shape modes past a bubble rising in water. Top: axisymmetric $(2,0)$ mode at $Bo=0.54$, with the real and imaginary parts shown in $(a)$ and $(b)$, respectively. Bottom: same for the asymmetric $(2,1)$ mode at $Bo=0.90$. For caption see figures \ref{fig:modesT05} and \ref{fig:rigidmodesWater}.}
 \label{fig:shapemodesWater}
\end{figure}
\subsubsection{Spatial structure of rigid and shape modes}
\label{structwater} 
Figure \ref{fig:rigidmodesWater} displays the structure of the three successive rigid modes encountered in the sequence analysed above. Subfigures $(a-b)$ look qualitatively similar to their counterparts in figure \ref{fig:modesT05}. However the `double-crescent' structure identified in the bottom part of the latter is not present here. The reason is that the Strouhal number of the primary low-frequency mode is approximately three times lower in water (see figure \ref{fig:St}). For this reason, the vertical distance separating vortices shed in the wake is much larger, so that the double-crescent structure closest to the bubble is located further downstream. Subfigures $(c)$ and $(e)$ compare the wake structure of the primary and secondary oscillatory rigid modes. They evidence the higher vortex shedding frequency of the latter, in line with the nearly twice as large dimensionless frequency noticed at point A' in figure \ref{fig:LinearStabilityWater} compared with that at point A. In the case of the stationary mode displayed in subfigure $(d)$, the azimuthal vorticity and velocity disturbances are seen to keep a constant sign all along the wake (the real part of the azimuthal velocity is null in this case, which is why the imaginary part is shown in the figure). This is the generic hallmark of wakes associated with stationary inclined paths of axisymmetric bodies; e.g. \cite{Tchoufag2014a}, \cite{Tchoufag2014b}. No vortex shedding takes place in that situation, the inclination of the body path resulting from the constant lift force generated by a pair of semi-infinite counter-rotating streamwise vortices. \\
\indent Figure \ref{fig:shapemodesWater} reveals the structure of the $(2,0)$ and $(2,1)$ shape modes at $Bo=0.54$ and $Bo=0.9$, respectively. The axial symmetry of the flow field is obvious in subfigures $(a-b)$, with a succession of toroidal eddies {\color{black}{of alternate signs}} shed downstream of the bubble. The red and blue contours evidence the fact that the bubble alternatively `inflates' and `deflates' along its axis while simultaneously shortening or lengthening in its equatorial plane. The streamlines in subfigures $(c-d)$ make the antisymmetry of the mode $(2,1)$ just as obvious. The wake is dominated by a series of double-crescent structures qualitatively similar to those identified in figure \ref{fig:modesT05}. The wavelength separating two consecutive structures is slightly shorter than that of the toroidal eddies in subfigures $(a-b)$, although the dimensionless frequency of the mode $(2,1)$ at $Bo=0.9$ is approximately $30\%$ lower than that of the mode $(2,0)$ at $Bo=0.54$ according to figure \ref{fig:LinearStabilityWater}$(b)$. However, these structures are advected downstream by the base flow, so that their wavelength is determined by the Strouhal number $St=\lambda_i/(2\pi U)$, not directly by $\lambda_i$. Therefore, what the comparison of the wavelengths for the two modes indicates is that the normalized rise speed $U$ decreases by slightly more than $30\%$ from $Bo=0.54$ to $Bo=0.9$, owing to the progressive flattening of the bubble (in the regime where bubbles keep an approximate oblate spheroidal shape, $U$ is known to vary approximately as $Bo^{-1/2}$ for $Bo\ll1$ and $Bo^{-1}$ for $Bo\gg1$; see equation (7-3) and the associated figure in \cite{Clift1978}).

\section{Respective magnitudes of rigid-body motions and time-dependent deformations}
\label{deform}
In the previous section, we identified `rigid' and `shape' modes qualitatively, stating that the former leave the bubble shape almost unchanged while the latter are associated with significant time-dependent shape oscillations. However, this distinction deserves a more quantitative basis. Moreover, determining how large the deformations at the threshold of path instability are, compared with the lateral drift of the bubble, is key to clarify the relative role of the various possible causes of instability present in the system. In order to quantify time-dependent deformations, we define a $(x,y,z)$ Cartesian coordinate system, the origin of which stands at the centroid of the bubble in the base state. We assume that the bubble performs planar zigzags in the vertical $(x,z)$ plane corresponding to the plane $\theta=(0,\pi)$ in the cylindrical $(r,\theta,z)$ system defined in \S\,\ref{numer}. At any position ${\bf{x}}_0$ on the undisturbed bubble-fluid interface, the normal displacement of the interface, $\hat\eta({\bf{x}}_0)$, may be decomposed into a rigid motion inducing a lateral (resp. vertical) displacement $\hat{T}_x$ (resp. $\hat{T}_z$) about the $x$ (resp. $z$) axis and an inclination $\hat{\psi}$ resulting from a rotation about the $y$ axis, augmented by a local volume-preserving deformation $\hat\zeta({\bf{x}}_0)$ in the normal direction. Projecting this decomposition onto the local normal unit vector ${\bf{n}}_0$, one has
\begin{equation}
\hat\eta({\bf{x}}_0)=\hat{T}_x({\bf{e}}_x\cdot{\bf{n}}_0)+\hat{T}_z({\bf{e}}_z\cdot{\bf{n}}_0)+\hat{\psi}({\bf{x}}_0\times{\bf{n}}_0)\cdot{\bf{e}}_y+\hat{\zeta}({\bf{x}}_0)\,,
\label{decomp}
\end{equation}
with ${\bf{e}}_x,{\bf{e}}_y,{\bf{e}}_z$ the unit vectors in the $x,y,z$ directions, respectively. Odd modes $m=\pm1$ do not induce any translation in the vertical direction, so that $\hat{T}_z=0$ for such modes. Conversely, the lateral displacement $\hat{T}_x$ and the inclination $\hat\psi$ of the bubble remain null for even modes $m=-2,0,2$. To determine the nonzero displacement and inclination associated with a given mode, we make use of a least-squares fitting technique. That is, we  define the local functional $\mathcal{F}_{\hat{T}_x,\hat{T}_z,\hat\psi}({\bf{x}}_0)\equiv\left\{\hat\eta-[\hat{T}_x({\bf{e}}_x\cdot{\bf{n}}_0)+\hat{T}_z({\bf{e}}_z\cdot{\bf{n}}_0)+\hat{\psi}({\bf{x}}_0\times{\bf{n}}_0)\cdot{\bf{e}}_y]\right\}^2$, integrate it over the entire bubble contour and minimize the corresponding residue with respect to the nonzero components of the triplet $(\hat{T}_x,\hat{T}_z,\hat\psi)$. {\color{black}{Given the linear framework of the present study, the magnitude of the displacements is arbitrary, and only their relative magnitudes with respect to a common reference makes sense. This is why}} we normalize the local displacement $\hat\eta({\bf{x}}_0)$ by $\hat T_x$ for odd modes and $\hat T_z$ for even modes, in order to make the translational displacement of the bubble centroid unity in all cases.\\
\begin{figure}
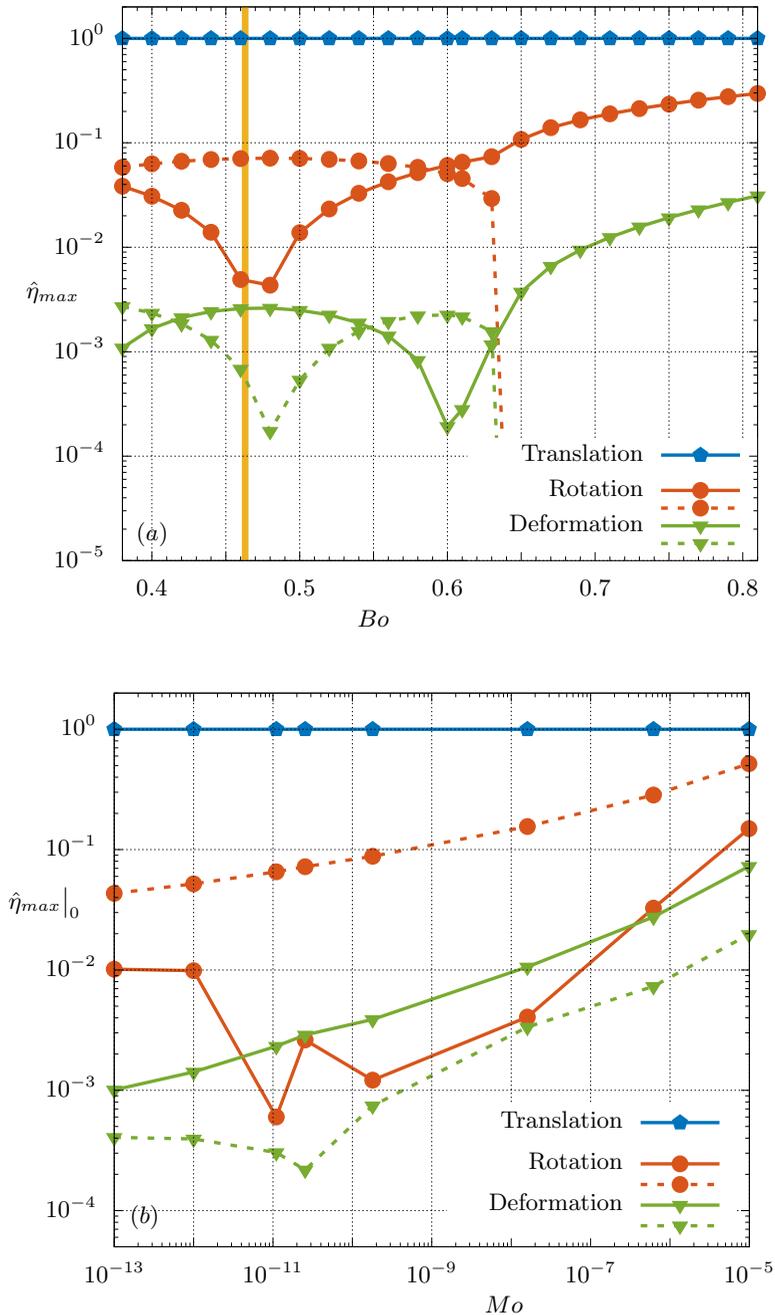

\vspace{5mm}
\centering
 {\input{Figures/Decomp_water.tex}\\
 \input{Figures/Decomp_neutral-curve.tex}
}
  \vspace{-109mm}\\
\hspace{-63mm}$(a)$
\vspace{86.5mm} \\
\hspace{-65mm}$(b)$
\vspace{13mm} \\
     \caption{Decomposition of the normal displacement of the bubble surface into a horizontal uniform translation, an inclination-induced displacement resulting from a rigid-body rotation, and a volume-preserving deformation. $(a)$: variation with $Bo$ for the most amplified rigid mode $|m|=1$ in the case of a bubble rising in water at $20^\circ\,$C; the yellow vertical line indicates the threshold of path instability; $(b)$: variation with $Mo$ at the path instability threshold. Each line figures the maximum of the corresponding component over the bubble contour; for each component, the solid and dashed lines refer to the real and imaginary parts, respectively. }
 \label{fig:decomp}
\end{figure}
\indent Thanks to this procedure, the evolution of the bubble shape and the relative magnitude of the interface displacements induced by the rigid-body rotation and volume-preserving deformations may be tracked along each branch of the bifurcation diagram. To save space, we do not discuss shape evolutions here (examples are provided by \cite{Bonnefis2019}), and focus on the relative magnitude of the various contributions and their physical origin. Figure \ref{fig:decomp}$(a)$ considers the case of the most amplified rigid mode for a bubble rising in water at $20^\circ\,$C{\color{black}{, i.e. the low-frequency oscillating mode up to $Bo\approx0.64$, followed by the upper branch of the stationary mode for higher $Bo$ (figure \ref{fig:LinearStabilityWater})}}. It shows how the maximum over the bubble contour of the three contributions to $\hat\eta$ vary with the Bond number. In the vicinity of the threshold ($Bo=0.463$), the inclination-induced displacement is seen to be essentially in quadrature with the lateral drift, which is the expected behaviour in a planar zigzagging motion if the bubble axis remains aligned with the path {\color{black}{(i.e. its inclination is maximum when the bubble reaches the centreline of its zigzagging path)}}. The relative magnitude of the maximum normal displacement resulting from this inclination is $7\times10^{-2}$ and experiences little variation up to $Bo\approx0.55$. At the threshold, the maximum of the volume-preserving deformation is almost in phase with the lateral drift. {\color{black}{This is because the rise speed reaches a maximum at each extremity of the zigzag \citep{Mougin2006}, resulting in some flattening of the bubble. Conversely, the rise speed is minimum when the bubble crosses the centreline of the zigzag, which slightly reduces its oblateness, yielding the small nonzero imaginary component of the deformation (dashed green line in figure \ref{fig:decomp}$(a)$).}} The relative magnitude of the deformation does not exceed $3\times10^{-3}$. This provides the confirmation that time-dependent deformations are extremely weak at the onset of path instability for bubbles rising in low-$Mo$ liquids. These findings also validate the `rigid' qualification employed for the primary low-frequency mode throughout \S\,\ref{unstable}. Conversely, the relative magnitude of deformations in modes $(2,0)$ and $(2,1)$ (not shown) is almost unity at their respective thresholds. The real (resp. imaginary) part of the displacements associated with the bubble inclination (resp. deformation) is typically one order of magnitude smaller than its imaginary (resp. real) counterpart. These weak components exhibit abrupt variations close to the threshold (and for one of them around $Bo=0.6$), presumably because such low-magnitude quantities are only determined with a limited accuracy by the least-squares technique. \\
\indent These two weak components increase sharply with $Bo$ and become dominant for $Bo\gtrsim0.55${\color{black}{, marking a change in the style of the bubble ascent}}. In particular, the real part of the displacement associated with the bubble inclination exceeds the imaginary part for $Bo\geq0.59$. This behaviour indicates that the bubble now exhibits a nonzero inclination at the extremities of its zigzagging path, so that its minor axis is no longer aligned with its velocity. {\color{black}{In between two successive extremities of the zigzag, the bubble now glides along its path rather than facing it. This makes the part of its surface pointing in the direction of the lateral drift slightly more pointed, while the opposite part becomes slightly more rounded. This is the reason why the imaginary part of the deformation is significant in this second style of zigzagging motion. Conversely, owing to the nonzero bubble inclination at the extremities of the zigzag, the rise speed is slightly lower than in the previous zigzag style. This mitigates the flattening of the bubble, thus reducing the real part of the volume-preserving deformation, as observed on the solid green line of figure \ref{fig:decomp}$(a)$ in the range $0.5\lesssim Bo\lesssim0.6$.}} \\
\indent At $Bo=0.64$, the imaginary parts of the rotation- and deformation-induced contributions fall to zero, a consequence of the change in nature of the primary rigid mode at the exceptional point revealed by figure \ref{fig:LinearStabilityWater}. {\color{black}{Beyond this point, the rigid mode is stationary, yielding a nonoscillatory inclined path. When the Bond number increases in the range $0.6-0.8$, the bubble gradually flattens and the size of the standing eddy at its back grows significantly, as panels $(h)$ and $(i)$ of figure \ref{fig:ContoursBaseFlow_DMST05} show. This makes the strength of the trailing vortices that set in at the path instability threshold increase with $Bo$, which translates directly into an increase of the lift force acting on the bubble, hence an increase of its inclination with the Bond number. This scenario is confirmed by the solid red line in figure \ref{fig:decomp}$(a)$, which shows that the maximum displacement associated with the bubble inclination increases by a factor of four from $Bo=0.64$ to $Bo=0.8$. The larger the bubble inclination, the lower its rise speed, which in turn makes its oblateness reduce compared to that in the base state. This is the origin of the sharp increase of the real part of the volume-preserving deformation (solid green line in figure \ref{fig:decomp}$(a)$), which gains two orders of magnitude between $Bo=0.6$ and $Bo=0.8$.}} Nevertheless, deformations remain small compared with the lateral drift, reaching only a relative magnitude of $3\times10^{-2}$ at $Bo=0.8$. \\
\indent Figure \ref{fig:decomp}$(b)$ shows how the relative magnitude of the three contributions to $\hat\eta$ vary with the Morton number at the path instability threshold, i.e. along the neutral curve of figure \ref{fig:neutralcurve}. {\color{black}{Since the corresponding unstable mode is always oscillatory, one expects the relative magnitude of the different contributions to behave qualitatively in the same way as in the case of water in the vicinity of the primary threshold.}} Indeed, in line with the behaviours noticed above, deformation-induced displacements are essentially in phase with the bubble lateral drift throughout the eight decades of $Mo$ explored here, while those associated with the bubble inclination are almost in quadrature. Again, we attribute the non-smooth variations of the weak real (resp. imaginary) component of the inclination-induced (resp. deformation-induced) contribution observed for $Mo\lesssim10^{-10}$ to the limitations of the minimization technique.  In contrast, the dominant component of these two contributions is seen to increase smoothly with the Morton number. The relative deformation-induced displacements, say $\hat\eta^{def}_{max}\big|_0$, follow the approximate power law $\hat\eta^{def}_{max}\big|_0\sim Mo^{0.23}$, increasing by two orders of magnitude from Galinstan ($\hat\eta^{def}_{max}\big|_0\approx10^{-3}$) to DMS-T11 ($\hat\eta^{def}_{max}\big|_0\approx7\times10^{-2}$). 
The relative inclination-induced displacements only grow by one order of magnitude, from $4\times10^{-2}$ to $5\times10^{-1}$, in between the same two limits. Nevertheless, the fact that these displacements reach $50\%$ of the lateral drift for DMS-T11 (to be compared with $7\%$ in the case of water) indicates that bubbles wobble significantly when their path becomes unstable in liquids having a Morton number four to five orders of magnitude higher than that of water.
\section{Summary and final discussion}
\label{conclu}
\subsection{Main findings}
\label{main}
Thanks to an innovative numerical approach allowing the linear stability of the viscous flow past freely rising and deforming bubbles to be assessed accurately, even when the rise Reynolds number is large, we explored the conditions under which the path of such bubbles deviates from a straight vertical line and the physical nature of the underlying instability modes over a wide range of properties of the carrying liquids. Present results extend over eight orders of magnitude of the Morton number. Over this range, the critical Bond number increases by a factor of $35$ from Galinstan to silicone oil DMS-T11 (yielding an increase by a factor of $2.6$ of the critical bubble size), whereas the critical Reynolds number decreases by a factor of $30$, from $\approx2100$ to $70$, in between the same two liquids. We showed that path instability always arises through the destabilization of a non-axisymmetric oscillatory mode associated with weak time-dependent changes of the bubble shape. Increasing the bubble size in a given fluid, the nature of this primary `rigid' mode stays unchanged if the Morton number is large enough, say $Mo\gtrsim10^{-7}$. Conversely, this mode gives birth to two stationary modes if the Morton number is low enough ($Mo\lesssim10^{-9}$). 
In the intermediate range $10^{-9}\lesssim Mo\lesssim 10^{-7}$, this mode remains oscillatory even far from the primary threshold but its growth rate varies non-monotonically with the bubble size, and it may even re-stabilise within a narrow size range. A second rigid mode, which turns to be stationary for intermediate and `high' Morton numbers but oscillatory for low Morton numbers also arises further away from the primary threshold. `Shape' modes distinct from the previous rigid modes in that they exhibit much larger oscillations of the bubble surface also become unstable as the bubble size is increased.\\ 
\indent Comparison of present predictions with reference experimental and numerical data revealed a very good agreement, both on the threshold (most often measured through the Bond or Galilei number, or alternatively the rise Reynolds number) and the reduced frequency of path oscillations (measured through a Strouhal number). This agreement fully supports the view that path instability of real gas bubbles results from a linear mechanism associated with a Hopf bifurcation. In contrast, oblique paths which are the hallmark of stationary modes have not been reported in experiments, nor in fully-resolved simulations. We see their absence as an indication that the changes introduced in the base flow by the primary non-axisymmetric oscillatory mode are significant enough to alter the next steps of the actual bifurcation sequence compared with the predictions of the linear approach. \\
\indent To better quantify time-dependent bubble deformations and their influence on the path instability mechanisms and characteristics, we employed two additional and complementary tools. First, a least-squares minimization technique allowed us to split the displacements of the bubble-fluid interface into contributions associated with a rigid-body motion and volume-preserving deformations. By doing so, we could among other things establish how the amplitude of the deformations varies with the Morton number at the onset of path instability. It turned out that the ratio of the maximum deformation to the maximum lateral drift of the bubble centroid is always much smaller than unity. In particular, this ratio is approximately $10^{-3}$ for Galinstan and $3\times10^{-3}$ for water. It exceeds $1\%$ only for liquids with $Mo\gtrsim1.5\times10^{-8}$, reaching $7\%$ for DMS-T11  {\color{black}{($Mo\!=\!9.9\times 10^{-6}$)}}. 
Second, we took advantage of the base states computed with the present numerical technique to update the predictions obtained in the framework of the frozen-shape approximation (FSA) by \cite{Cano2016}. With this procedure, any difference in the nature of the primary bifurcation or/and the corresponding threshold may be ascribed to effects of time-dependent bubble deformations. We found that, despite their weak magnitude, these deformations lower significantly the threshold whatever the Morton number. Hence, a general conclusion of the present study is that deformations always promote path instability in the linear framework. Differences between the predicted critical size of deformable and non-deformable bubbles increase gradually from $8\%$ at $Mo\approx10^{-13}$ to $23\%$ at $Mo\approx6\times10^{-7}$, before decreasing sharply to nearly $15\%$ for larger $Mo$. For $Mo\lesssim7\times10^{-11}$, FSA correctly predicts that the most unstable mode is a `low-frequency' one with $St=\mathcal{O}(0.01)$ at the threshold. Similarly, it rightly predicts that path instability arises through a `high-frequency' mode with $St=\mathcal{O}(0.1)$ at the threshold for $Mo\gtrsim3\times10^{-7}$. The predicted reduced frequencies lie $20$ to $30\%$ below those obtained with deformable bubbles, but their variations with the Morton number are qualitatively similar. In contrast, FSA dramatically fails to detect that the most unstable mode is still oscillatory in the intermediate Morton number range $7\times10^{-11}\lesssim Mo\lesssim3\times10^{-7}$. 
\subsection{Discussion}
\label{discu}
To better understand the above issue and gain some insight into the subtle balances involved in the various regimes encountered by varying the fluid properties, it is useful to put the first stages of the bifurcation diagrams obtained with deformable bubbles in perspective with those determined by \cite{Tchoufag2014b} who assumed a perfectly oblate spheroidal bubble with a frozen shape. By increasing the geometrical aspect ratio $\chi$ in the range $[2.21,2.30]$ they identified three successive phenomenologies summarized in their figures 1 and 2. 
For $2.21\leq\chi\leq2.23$, the system first becomes unstable through a low-frequency mode which, by increasing the bubble size (i.e. the Reynolds number in their case), turns into a pair of stationary modes, one of which is quickly damped as the bubble size is increased further while the other gets strongly amplified (their figure 1). Present findings summarized in figure \ref{fig:LinearStabilityWater} indicate that this is exactly what happens with deformable bubbles rising in water and more generally in low-$Mo$ liquids. That the scenario relevant to frozen spheroidal bubbles remains unchanged in the case of freely-deforming bubbles rising in low-$Mo$ liquids strongly supports the view that time-dependent deformations do not play any causal role in the mechanisms governing the first stages of path instability in such fluids, although they lower the critical bubble size by $18\%$ for water and increase the reduced frequency by nearly $30\%$, as may be deduced from the two neutral curves in figures \ref{fig:neutralcurve} and \ref{fig:St} for $Mo=2.54\times10^{-11}$. The wake alone is not the source of the instability, since it is still stable at the FSA threshold \citep{Cano2016}. Therefore, as already stated by \cite{Bonnefis2023} in the case of water, the key explanation of path instability in low-Morton-number fluids lies in the specific physical processes accounted for in FSA, namely the degrees of freedom allowed by the rigid-body buoyancy-driven motion of the bubble, and the back reaction imposed by the fluid to this motion by the constant-force and zero-torque constraints. \\
\indent Increasing the bubble aspect ratio to the range $2.23\leq\chi\leq2.25$, \cite{Tchoufag2014b} observed that, although it is still present, the low-frequency mode is no longer unstable, its growth rate keeping weak negative values. In the approximation they used, path instability then arises through a stationary bifurcation, similar to what the FSA neutral curves  (pale grey lines) in figures \ref{fig:neutralcurve} and \ref{fig:St} indicate for $7\times10^{-11}\lesssim Mo\lesssim3\times10^{-7}$. The corresponding stationary mode is that associated with the wake instability observed when the bubble is held fixed in a uniform stream. Hence, one has to conclude that the additional degrees of freedom offered by the free motion of a frozen-shape bubble are not sufficient to change qualitatively the overall dynamics of the system in that range. 
The yellow neutral curve $St=f(Mo)$ in figure \ref{fig:St} proves that, despite their weak relative magnitude (from $0.4\%$ for DMS-T00 to $2\%$ for $Mo=3\times10^{-7}$ according to figure \ref{fig:decomp}$(b)$), deformations are able to restore the instability of the oscillatory mode and to make it more unstable than the stationary mode, keeping the near-threshold path instability phenomenology unchanged with respect to the one prevailing in the low-$Mo$-range. Nevertheless, expanding the growth rate in the form $\lambda_r\approx\alpha(Mo)(Bo-Bo_c(Mo))$, with $Bo_c$ the threshold Bond number, and considering variations of $\alpha(Mo)$ from water (figure \ref{fig:LinearStabilityWater}$(a)$) to DMS-T02 (figure \ref{fig:LinearStabilityDMST02}$(a)$) \textit{via} DMS-T00 (not shown), one finds $\alpha(Mo)\sim Mo^{-0.15}$. Hence, the larger $Mo$ the slower the increase of the growth rate of the oscillatory mode with the distance to the threshold, despite the fact that the magnitude of surface deformations at the threshold increases as $Mo^{0.23}$ (figure \ref{fig:decomp}$(b)$). This confirms that the damping mechanism predicted by FSA, be it with the schematic spheroidal shape prescribed by \cite{Tchoufag2014b} or with the more realistic shape corresponding to the actual base state, is still active in this intermediate $Mo$-range in the case of deformable bubbles. The reasons for this damping are not obvious. For a non-deforming body, the eigenvalues of the system depend directly on the drag and torque associated with a small edgewise translation or a slow rotation about the major axis perpendicular to the plane of motion \citep{Fabre2011}. We believe that when the bubble geometrical anisotropy is varied, small changes in the structure of the base flow, especially in the near wake, make these loads vary in a non-monotonic manner with $\chi$, hence with $Mo$, which makes eventually the real part of the eigenvalues associated with the oscillatory mode be positive in some ranges of the parameter space but slightly negative in others. \\
\indent Last, for $\chi\geq2.25$, \cite{Tchoufag2014b} found the first unstable mode to be oscillatory again. Similar to what is noticed on the FSA neutral curve in figure \ref{fig:St}, the corresponding oscillations are much faster than those observed for $2.21\leq\chi\leq2.23$, the two oscillatory modes predicted by FSA in the two subdomains being distinct (pale and dark grey lines in figures \ref{fig:neutralcurve} and \ref{fig:St}). Deformation effects totally smooth out this abrupt change by allowing the frequency to increase gradually throughout the intermediate $Mo$-range discussed above. Increasing the Reynolds number at a given $\chi$, i.e. the bubble size in a given fluid, \cite{Tchoufag2014b} found that the next mode that becomes unstable is the stationary mode associated with wake instability. The FSA prediction obtained with the actual base flow confirms this conclusion. For instance, the dark grey line in figure \ref{fig:neutralcurve} shows that, for DMS-T05, the `high-frequency' mode becomes unstable at $Bo\approx5.0$. Then the stationary mode (pale grey line) becomes unstable in turn at $Bo=5.85$, i.e. for a bubble only $8\%$ larger than at the primary threshold. This sequence subsists when the bubble deforms freely, but the gap between the two thresholds increases considerably, with the oscillatory and stationary modes becoming unstable at $Bo=3.4$ and $Bo=5.95$, respectively. Moreover, the influence of deformation manifests itself through the emergence of the $(2,0)$ axisymmetric shape mode that becomes unstable in between these two thresholds. 
\indent The above comparisons establish that the role of bubble deformations on the most unstable non-axisymmetric mode responsible for path instability deeply differs among the three $Mo$-ranges identified by FSA. This role is limited to mere, albeit significant, quantitative changes in the critical bubble size and oscillations frequency in the low- and `high'-$Mo$ regimes, whereas it induces a complete change in the nature of the most unstable mode in the intermediate-$Mo$ regime.\\
\indent  A weakly nonlinear approach inspired by those developed by \cite{Fabre2012} and \cite{Tchoufag2015} for buoyancy/gravity-driven rigid bodies is desirable to extend present predictions to slightly supercritical bubble sizes. By incorporating nonlinear corrections to the base state resulting from the primary non-axisymmetric unstable mode, this extension may in particular allow the disappearance of the stationary mode at any Morton number to be rationalized. {\color{black}{Such a weakly nonlinear approach would also settle the question of the supercritical or subcritical nature of the primary bifurcation in each Morton number range, which the present linear approach cannot answer. }}Another  more modest but much cheaper approach worth exploring is the extension to bubbles (and more generally to rigid bodies with a moderate or low body-to-fluid density ratio) of the `aerodynamic' reduced-order model derived by \cite{Fabre2011} for heavy freely-falling two-dimensional bodies. This model should be able to predict accurately the most unstable or least stable eigenvalues of the system for bubbles with a frozen shape, provided that the drag and torque coefficients associated with a small edgewise translation and a slow rotation of the bubble are computed in the relevant base flow. This is the route to be followed to quantify the influence of base flow variations resulting from slight changes in the interface geometry on the hydrodynamic reaction experienced by bubbles close to the threshold of path instability.    \\

{\bf Declaration of Interests}. The authors report no conflict of interest.\\

{\bf Author ORCIDs}.\\
P. Bonnefis, https://orcid.org/0000-0003-2825-2974; \\
J. Sierra-Ausin, https://orcid.org/0000-0001-6036-5093; \\
D. Fabre, https://orcid.org/0000-0003-1875-6386; \\
J. Magnaudet, https://orcid.org/0000-0002-6166-4877.
\bibliographystyle{jfm}
\bibliography{bubblestab}

\end{document}